\documentclass[manuscript]{aastex}
\usepackage{emulateapj5}

\shorttitle{CHANDRA X-RAY OBSERVATIONS OF NGC~4697}
\shortauthors{SARAZIN, IRWIN, \& BREGMAN}

\slugcomment{The Astrophysical Journal, in press}

\begin{document}

\title{Chandra X-ray Observations of the X-ray Faint Elliptical Galaxy
NGC~4697}

\author{Craig L. Sarazin\altaffilmark{1},
Jimmy A. Irwin\altaffilmark{2,3},
and Joel N. Bregman\altaffilmark{2}}

\altaffiltext{1}{Department of Astronomy, University of Virginia,
P. O. Box 3818, Charlottesville, VA 22903-0818;
sarazin@virginia.edu}

\altaffiltext{2}{Department of Astronomy, University of Michigan,
Ann Arbor, MI 48109-1090; jirwin@astro.lsa.umich.edu, jbregman@umich.edu}

\altaffiltext{3}{Chandra Fellow}

\begin{abstract}
A {\it Chandra} ACIS S3 observation of the X-ray faint
elliptical galaxy NGC~4697 resolves much of the X-ray emission
(61\% of the counts from within one effective radius) into 90 point sources,
of which $\sim$80 are low mass X-ray binaries (LMXBs)
associated with this galaxy.
The dominance of LMXBs indicates that X-ray faint early-type galaxies
have lost much of their interstellar gas.
On the other hand, a modest portion of the X-ray emission from
NGC~4697 is due to hot gas.
Of the unresolved emission, it is likely that about half is
from fainter unresolved LMXBs, while the other half
($\sim$23\% of the total count rate) is from interstellar gas.
The X-ray emitting gas in NGC~4697 has a rather low temperature
($kT = 0.29$ keV).
The emission from the gas is very extended, with a much flatter
surface brightness profile than the optical light,
and has an irregular, L-shaped morphology.
The physical state of the hot gas is uncertain;
the X-ray luminosity and extended surface brightness are inconsistent
with a global supersonic wind, a partial wind, or a global cooling inflow.
The gas may be undergoing subsonic inflation, rotationally induced outflow,
or ram pressure stripping.
X-ray spectra of the resolved sources and diffuse emission show that
the soft X-ray spectral component, found in this and other X-ray
faint ellipticals with {\it ROSAT}, is due to interstellar gas.
The cumulative LMXB spectrum is well-fit by thermal bremsstrahlung
at $kT = 8.1$ keV, without a significant soft component.

NGC~4697 has a central X-ray source with a luminosity of
$L_X = 8 \times 10^{38}$ ergs s$^{-1}$, which may be due to
an AGN and/or one or more LMXBs.
At most, the massive black hole at the center of this galaxy is radiating
at a very small fraction ($\le 4 \times 10^{-8}$)
of its Eddington luminosity.

Three of the resolved sources in NGC~4697 are supersoft sources.
In the outer regions of NGC~4697, seven of the LMXBs (about 20\%)
are coincident with candidate globular clusters,
which indicates that globulars have a high probability of containing
X-ray binaries compared to the normal stellar population.
We suggest that all of the LMXBs may have been formed in globulars.
The X-ray--to--optical luminosity ratio for the LMXBs in NGC~4697 is
$L_X$(LMXB, 0.3--10 keV)$/L_B = 8.1 \times 10^{29}$ ergs s$^{-1}$
$L_{B\odot}^{-1}$,
which is about 35\% higher than the value for the bulge of M31.
Other comparisons suggest that there are significant variations
(factor of $\ga$2) in the LMXB X-ray--to--optical ratios of early-type
galaxies and spiral bulges.
The X-ray luminosity function of NGC~4697 is also flatter than that
found for the bulge of M31.
The X-ray luminosities (0.3--10 keV) of the resolved LMXBs range from
$\sim$$5 \times 10^{37}$ to $\sim$$2.5 \times 10^{39}$ ergs s$^{-1}$.
The luminosity function of the LMXBs has a ``knee'' at
$3.2 \times 10^{38}$ ergs s$^{-1}$, which is approximately the Eddington
luminosity of a 1.4 $M_\odot$ neutron star (NS).
This knee appears to be a characteristic feature of the LMXB population
of early-type galaxies,
and we argue that it separates black hole and NS binaries.
This characteristic luminosity could be used as a distance estimator.
If they are Eddington limited, the brightest LMXBs contain fairly massive
accreting black holes.
The presence of this large population of NS and massive BH stellar remnants
in this elliptical galaxy shows that it (or its progenitors) once contained
a large population of massive main sequence stars.
\end{abstract}

\keywords{
binaries: close ---
galaxies: elliptical and lenticular ---
galaxies: ISM ---
X-rays: galaxies ---
X-rays: ISM ---
X-rays: stars
}

\section{Introduction} \label{sec:intro}

X-ray observations, starting with the {\it Einstein} X-ray Observatory,
have shown that elliptical and S0 galaxies are luminous sources of X-ray
emission
(e.g., Forman, Jones, \& Tucker 1985).
At least for the X-ray luminous early-type galaxies (defined as
those having a relatively high ratio of X-ray to optical luminosity
$L_X/L_B$),
it is clear that
the bulk of the X-ray luminosity is from hot interstellar
hot ($\sim$$10^7$ K) interstellar gas
(e.g., Forman et al.\ 1985;
Trinchieri, Fabbiano, \& Canizares 1986).
Presumably, the material comprising the interstellar medium
(ISM) originated from stellar mass loss in the galaxy.
There is a strong correlation between the X-ray and blue luminosities
of early-type galaxies $L_X \propto L_B^{1.7-3.0}$
(Canizares, Fabbiano, \& Trinchieri 1987;
White \& Davis 1997; Brown \& Bregman 1998).
However, there is a large dispersion in the X-ray luminosities of
early-type galaxies of a given optical luminosity.
Two galaxies with similar blue luminosities might have X-ray luminosities
that differ by as much as a factor of 100
(Canizares et al.\ 1987; Fabbiano, Kim, \& Trinchieri 1992;
Brown \& Bregman 1998).
We will refer to galaxies which have a very low $L_X/L_B$ ratio
as ``X-ray faint.''
In these X-ray faint ellipticals, much of the hot interstellar gas
may have been lost in galactic winds
(Loewenstein \& Mathews 1987; David, Forman, \& Jones 1991),
or by ram pressure stripping by ambient intracluster or intragroup gas
(White \& Sarazin 1991).
The source of the bulk of the X-ray emission in X-ray faint galaxies
is uncertain;
it might be due to low-mass X-ray binaries (LMXBs) like those seen in
the bulge of our Galaxy
(e.g., White, Nagase, \& Parmar 1995),
or to an active galactic nucleus (AGN)
(Allen, di Matteo, \& Fabian 2000),
or to interstellar gas
(Pellegrini \& Fabbiano 1994),
or to fainter stellar sources such as
active M stars or RS CVn binaries
(e.g., Pellegrini 1994).

In general, X-ray faint galaxies exhibit significantly different
X-ray spectral properties than their X-ray bright counterparts.
The X-ray bright galaxies are dominated by thermal emission at
$kT \sim 0.8$ keV due to their hot interstellar medium (ISM).
On the other hand, the X-ray faint galaxies exhibit two distinct spectral
components:
first, they have a hard $\sim$5--10 keV component,
most easily seen in {\it ASCA} spectra
(Matsumoto et al.\ 1997),
which is roughly proportional to the optical luminosity of the galaxy.
Actually, 
both X-ray faint and X-ray bright early-type galaxies appear to have
this hard X-ray component, 
which is roughly proportional to the optical luminosity of the galaxy.
This suggests that the hard component is due to low-mass X-ray binaries
(LMXBs) like those seen in the bulge of our Galaxy
(e.g., White, Nagase, \& Parmar 1995).
In some cases, an AGN may contribute to the hard component
(Allen et al.\ 2000),
or it might be due to hot gas.
However, the ASCA observations do not resolve this component into
discrete sources, nor to they provide much detailed information on its
spectrum.

X-ray faint galaxies also have
a very soft ($\sim$0.2 keV) component, whose origin is uncertain
(Fabbiano, Kim, \& Trinchieri 1994; Pellegrini 1994; Kim et al.\ 1996).
This difference in X-ray spectral characteristics is seen in the individual
X-ray spectra of a number of X-ray bright and faint galaxies, and in the
X-ray ``colors'' or hardness ratios determined for larger samples
(Irwin \& Sarazin 1998b).
Suggested stellar sources for the very soft emission in X-ray faint ellipticals
include active M stars, RS CVn binaries, or supersoft sources, but none of
these appears to work quantitatively
(Pellegrini \& Fabbiano 1994; Irwin \& Sarazin 1998a).
It is possible that the soft X-rays are due to warm (0.2 keV) ISM
(Pellegrini \& Fabbiano 1994).
Recently, we proposed that the very soft emission in X-ray
faint ellipticals might be due to the same LMXBs responsible for the
hard emission
(Irwin \& Sarazin 1998a,b).
Little is known about the very soft X-ray properties of Galactic LMXBs since
most lie in directions of high Galactic hydrogen
column densities, so their soft X-ray emission is heavily absorbed.
However, the origin of the very soft component in X-ray faint ellipticals
remains something of a mystery.

With the superb spatial resolution of the {\it Chandra} X-ray
Observatory, it should be possible to resolve the emission from
nearby X-ray faint early-type galaxies into LMXBs, if this is
indeed the source of their emission.
The clear test of the origin of the X-ray emission from X-ray faint
ellipticals is to see if it resolves into discrete LMXBs.
If the hard and/or soft X-ray emission from X-ray faint galaxies is due
to LMXBs, it should come from a relatively small number of individually
bright sources.
Recently, we used a deep {\it ROSAT} HRI observation to detect a
number of discrete X-ray sources in the nearby, optically luminous,
X-ray faint elliptical NGC~4697
(Irwin, Sarazin, \& Bregman 2000, hereafter ISB).
However, the bulk of the X-ray emission was not resolved.
In ISB, we also simulated a 40 ksec {\it Chandra} observation of NGC~4697,
and showed that it should be possible to detect $\sim$100 LMXBs
if they provide the bulk of the emission.
Here, we present the results of exactly this observation.
At a distance of 15.9 Mpc
(Faber et. al.\ 1989; assuming a Hubble constant
of 50 km s$^{-1}$ Mpc$^{-1}$),
NGC~4697 is among the closest optically luminous, X-ray
faint early-type galaxies.
Given its proximity, NGC~4697 is an ideal target for detecting
the LMXB population.
It is sufficiently X-ray faint that diffuse ISM emission should not 
bury the emission of the fainter LMXBs.
It should be possible to detect the LMXB population down to a level
a luminosity of $\sim 5 \times 10^{37}$ ergs s$^{-1}$.
The purposes of the {\it Chandra} observation are
to resolve and study the LMXB population of NGC~4697,
to determine the source of both the hard and soft spectral components,
and
to detect or place strong limits on any residual diffuse (possibly
gaseous) emission.

Some initial results of this observations were presented in
Sarazin, Irwin, \& Bregman (2000; hereafter Paper I).
In the present paper, we give the detailed properties of the sources.
In \S~\ref{sec:obs}, the observation and data analysis are described.
The overall X-ray image is discussed in \S~\ref{sec:image}.
The properties of the resolved sources are given in \S~\ref{sec:src},
while the remaining unresolved emission is analyzed in \S~\ref{sec:diffuse}.
In \S~\ref{sec:spectra}, the X-ray spectral properties of the galaxy are
derived.
The interpretation of the results is discussed in \S~\ref{sec:discuss},
and our conclusions are summarized in \S~\ref{sec:conclude}.

\section{Observation and Data Reduction} \label{sec:obs}

NGC~4697 was observed on 2000 January 15-16
with the ACIS-23678 chips operated at a temperature of -110 C and with
a frame time of 3.2 s.
The pointing was determined so that the entire galaxy was located on the S3
chip and so that the center of the galaxy was not on a node boundary of the
chip.
Although a number of serendipitous sources are seen on the other chips,
the analysis of NGC 4697 in this paper will be based on data from the
S3 chip alone.
The total exposure for the S3 chip was 39,434 s.
The data were telemetered in Faint mode, and
only events with ASCA grades of 0,2,3,4, and 6 were included.
We excluded bad pixels, bad columns, and the columns next to
bad columns and to the chip node boundaries.
We checked for periods of incorrect aspect solution, and none were found.
The ACIS S3 chip is known to experience occasional periods of high background
(Markevitch 2000a;
Markevitch et al.\ 2000).
We searched for such background flares by determining the lightcurve of
the total count rate in the S3 chip binned in 40 s intervals;
the total count rate is mainly due to background.
No background flares were found, and the background rate was constant
at about 0.9 s$^{-1}$, which was the quiescent rate for the S3 chip
at the time of the observation.
Only events with photon energies in the range 0.3 to 10 keV were included
in our analysis.

This observation was processed at a time when the standard pipeline
processing introduced a boresight error of about 8\arcsec\ in the
absolute positions of X-ray sources.
We corrected for this using optical identifications and
positions of X-ray sources.
As noted below (\S~\ref{sec:src}), a number of the X-ray sources have
faint optical identifications.
Of these, three had accurate positions given in the USNO-A2.0 optical
catalog
(Monet et al.\ 1998).
The optical and X-ray positions were offset by essentially the same
amount and in the same direction (to within 0.3\arcsec).
We applied the average of these offsets to the X-ray positions.
We believe that the quoted absolute positions are accurate
to about 0.5\arcsec\ near the center of the S3 image, with larger errors
further out.

Determining the background for these observations proved to be difficult.
We initially tried to use background from a series of nearly ``blank sky''
observations compiled by Markevitch (2000a,b).
For the S3 chip, most of the exposure in these background files occurs
at high Galactic latitude, mainly from the deep survey field of
Mushotzky et al.\ (2000),
These observations were taken when the focal plane temperature was -110 C
(as for our observations), and were screened to remove background flares;
thus, they should be appropriate for our data.
The blank-sky background file gave a surface brightness in the hard
(2 - 10 keV) band which agreed well with the hard surface brightness
in our data in outer parts of the S3 chip away from NGC~4697.
However, our data had significantly more soft emission in the outer parts.
In the outermost parts of the S3 chip,
this extra soft emission had a fairly uniform surface brightness.
While part of this emission is due to an extended soft component of
NGC~4697 (\S~\ref{sec:diffuse_spatial}),
the constant soft surface brightness
component appeared to be an additional background component.
We also determined the background for the region of the I3 chip which
is furthest from NGC~4697 and along its minor axis.
Our data also show a soft X-ray excess in this region relative to blank
sky fields.
We also examined the long {\it ROSAT} PSPC pointed observation of the galaxy
(ISB),
which showed excess soft emission in the outer parts of the field far
from NGC~4697 or other sources.

On the sky, NGC~4697 is located near the edge of the North Polar Spur (NPS),
a soft X-ray Galactic feature
(e.g., Snowden et al.\ 1997).
The {\it ROSAT} All Sky Survey (RASS) images of this region
(e.g., Snowden et al.\ 1997) show a strong
soft X-ray excess at the position of NGC~4697, which is consistent
with the soft excess seen in the outer parts of the pointed  PSPC
observation.
The excess is most prominent in the R4R5 {\it ROSAT} band, which
corresponds roughly to 0.4--1 keV.
The excess in the {\it Chandra} data is mainly in our soft band
(0.3--1 keV).
The level of the soft excess due to the NPS is such that it would
significantly affect the background in our observation.

We used the surface brightness profile of NGC~4697 in various energy
bands in both the {\it ROSAT} PSPC observation and in our {\it Chandra}
data to determine the portion of the extended soft emission which
was due to the NPS and which was due to NGC~4697.
Simple thermal models with temperatures of $\sim 2 \times 10^6$ K 
and a variety of abundances were used to convert the count rates for
the soft excess between {\it ROSAT} and {\it Chandra}.
In the outer parts of the S3 chip, we found that 55\% of the soft excess
was due to the NPS, while 45\% was due to NGC~4697.
For studying the properties of extended diffuse emission, we determine
the background by combining the blank sky background of Markevitch
(2000a,b) with background determined from the outer portion of the S3
chip in our observation, and weighted by 45\% and 55\%, respectively.
We added a systematic error to the background statistical error based
on the difference in these two determination of the background.

In determining the properties of the resolved sources (\S~\ref{sec:src}),
background for each source was determined from a region around the
source.
This background properly includes all sources of diffuse emission, including
the diffuse emission from NGC~4697.
This also avoids the problem of the effect of the NPS as discussed above.

All of the X-ray spectra were extracted using the PI values for the
events in order to correct for gain variations over the S3 chip.
The PI values were recomputed using the gain file
acisD1999-09-16gainN0004.fits,
appropriate for operating temperature of -110 C and frametime of 3.2 s.
The spectral responses were based on the FITS Embedded Function (FEF) files
FP-110\_D1999-09-16fef\_piN0002.
All of the spectra discussed here are extracted from extended regions,
which cover many of the 32$\times$32 pixel regions covered by individual FEFs.
We used the {\sc runextrspec} software package kindly provided by Alexey
Vikhlinin
(Vikhlinin, Markevitch, \& Murray 2001),
to extract spectra and to determine the response matrices.
This program weights the response files by the number of counts in each
region.

%
% Figure xray_cen
%
\centerline{\null}
\vskip2.75truein
\includegraphics{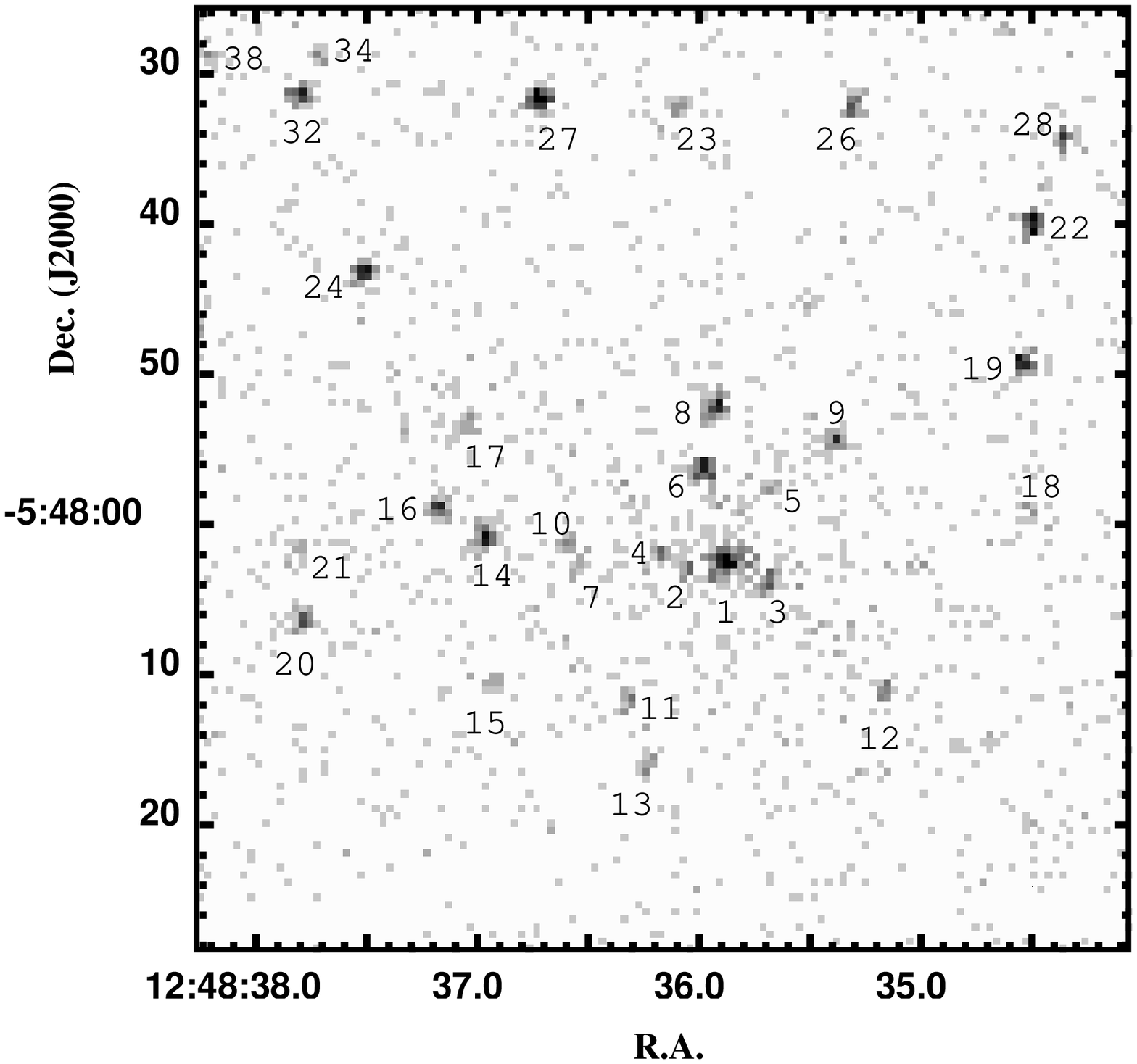}
\figcaption{
The {\it Chandra} S3 image of an approximately 1\arcmin$\times$1\arcmin\ 
region near the center of NGC~4697.
The greyscale varies with the square root of the X-ray surface brightness,
which ranges from 1 to 25 counts per pixel.
(The ACIS pixels are 0\farcs492 square.)
The positions of detected sources in the image are indicated by their
source numbers in Table~\protect\ref{tab:src};
the source numbers are ordered by increasing distance from the center of
the galaxy.
Src.~1 is located at the optical center of the galaxy.
\label{fig:xray_cen}}

\vskip0.2truein

\section{X-ray Image} \label{sec:image}

The raw {\it Chandra} S3 chip X-ray image was shown in
Figure~1 of Paper I.
This image showed the basic result of the
{\it Chandra} observation:
much of the emission from the galaxy is resolved into individual point
sources of X-rays.
The raw X-ray image of a 1\arcmin$\times$1\arcmin\ central region of the
NGC~4697 is shown in Figure~\ref{fig:xray_cen}.
The greyscale is proportional to the square root of the X-ray surface
brightness, and the values vary between 1 and 25 cts pixel$^{-1}$.
The numbers identify the detected sources in this regions as
listed in Table~\ref{tab:src} below.
The optical center of the galaxy coincides with the position of
Src.~1.
The center of the image has been offset from the optical because there
are more sources to the ENE than WSW of the galaxy center.
The density of resolved sources near the center of the galaxy is high,
and they may become confused in this region.

%
% Figure xray_smo
%
\centerline{\null}
\vskip2.65truein
\includegraphics{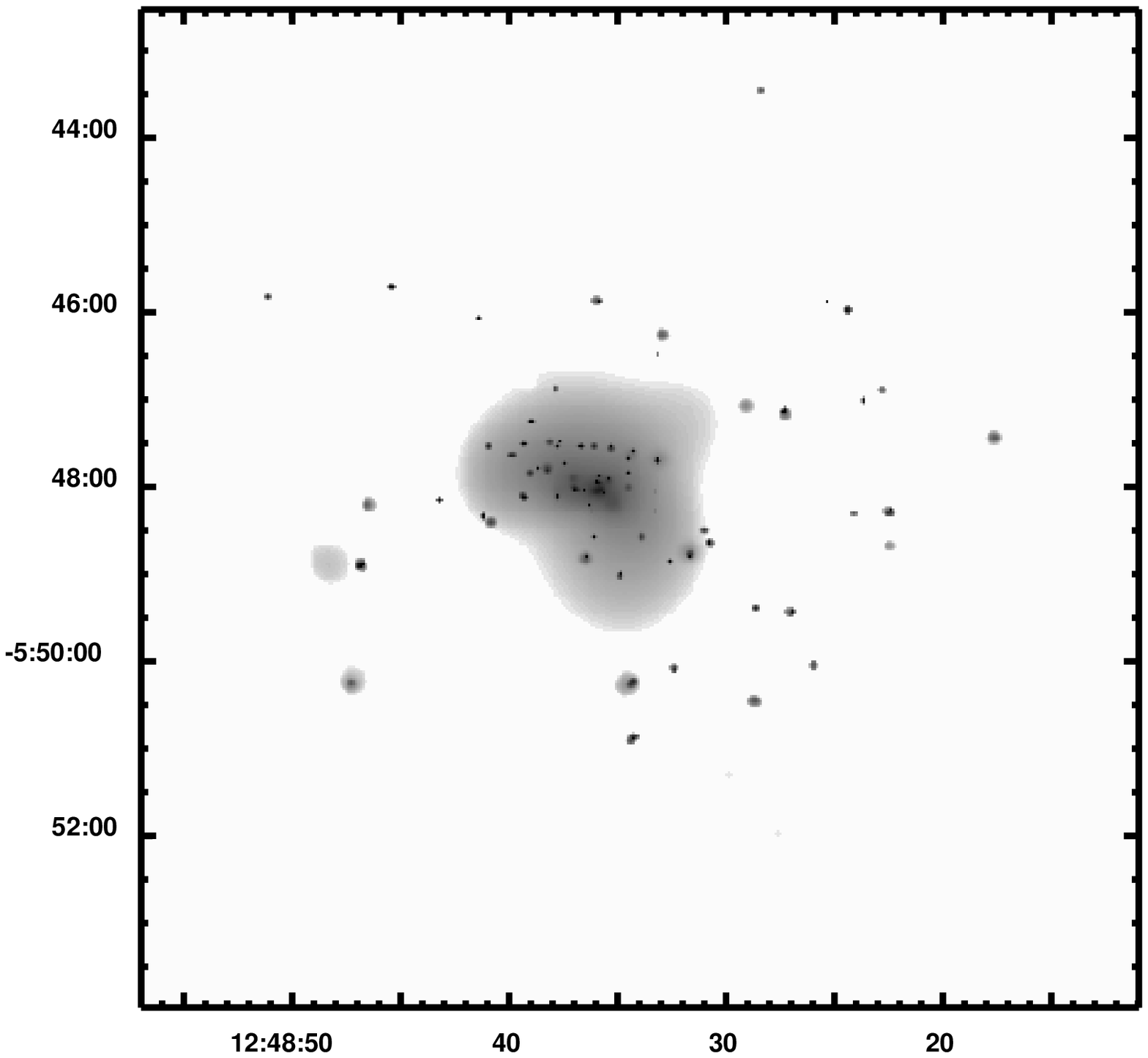}
\figcaption{
An adaptively smoothed version of the entire {\it Chandra} S3 X-ray image
of the region around NGC~4697 in the 0.3--10 keV band.
The image was adaptively smoothed to a signal-to-noise ratio of 3 per
smoothing beam, and corrected for exposure and background.
Regions with an exposure of less than 20 ksec were removed to avoid
artifacts at the chip edges.
The grey scale is logarithmic, and ranges from $2 \times 10^{-7}$ to
$10^{-3}$ cts pixel$^{-1}$ s$^{-1}$.
(The ACIS pixels are 0\farcs492 square.)
\label{fig:xray_smo}}

\vskip0.2truein

In order to image the fainter, more diffuse emission, we adaptively
smoothed the entire {\it Chandra} S3 X-ray image
to a minimum signal-to-noise ratio of 3 per smoothing beam.
The image was corrected for exposure and background.
Regions with an exposure of less than 20 ksec were removed to avoid
artifacts at the chip edges.
This adaptively smoothed image is shown in Figure~\ref{fig:xray_smo}.
The grey scale is logarithmic and 
covers surface brightnesses which range over a factor of $\ga$4000.
The central 4\arcmin$\times$4\arcmin\ region of the adaptively smoothed
image was presented as Figure~2 in Paper I, showing only the brighter
regions.

For comparison, Figure~\ref{fig:opt} shows the Digital Sky Survey
optical image of the same region as shown in Figure~\ref{fig:xray_smo}.
The circles indicate the positions of individual X-ray sources which
are identified in \S~\ref{sec:src} and Table~\ref{tab:src} below.
The distribution of X-ray point sources in
Figure~1 of Paper I
appears to be elongated at the same position angle as the optical image
of this E6 galaxy
(67$^\circ$;
Jedrzejewski, Davies, \& Illingworth 1987;
Faber et al.\ 1989;
Peletier et al.\ 1990);
this is discussed in detail in \S~\ref{sec:src_distr}.
(The small background galaxy LCRS B124537.4-053024, which is located
at the right side of Figure~\ref{fig:opt}, is beyond the edge of the S3 chip.)

Figures~\ref{fig:xray_cen}, \ref{fig:xray_smo}, and Figure~1 of Paper I
show that the
X-ray emission from NGC~4697 is mainly due to resolved sources,
which are discussed in more detail in \S~\ref{sec:src} below.
However, there is also some spatially extended, unresolved emission.
Near the center of the image (Figure~2, Paper I), the unresolved emission
follows the elliptical distribution of optical light in the galaxy.
However, at larger radii the unresolved emission becomes somewhat
irregular, and has something of an ``L-shaped" distribution, with extension
to the east and south.
The unresolved emission also is more extended than the optical light of
the galaxy at the same relative surface brightness levels, particularly
along the optical minor axis of the galaxy.
The spatial distribution of the unresolved emission is discussed in
more detail in \S~\ref{sec:diffuse_spatial},

%
% Figure opt
%
\centerline{\null}
\vskip2.75truein
\includegraphics{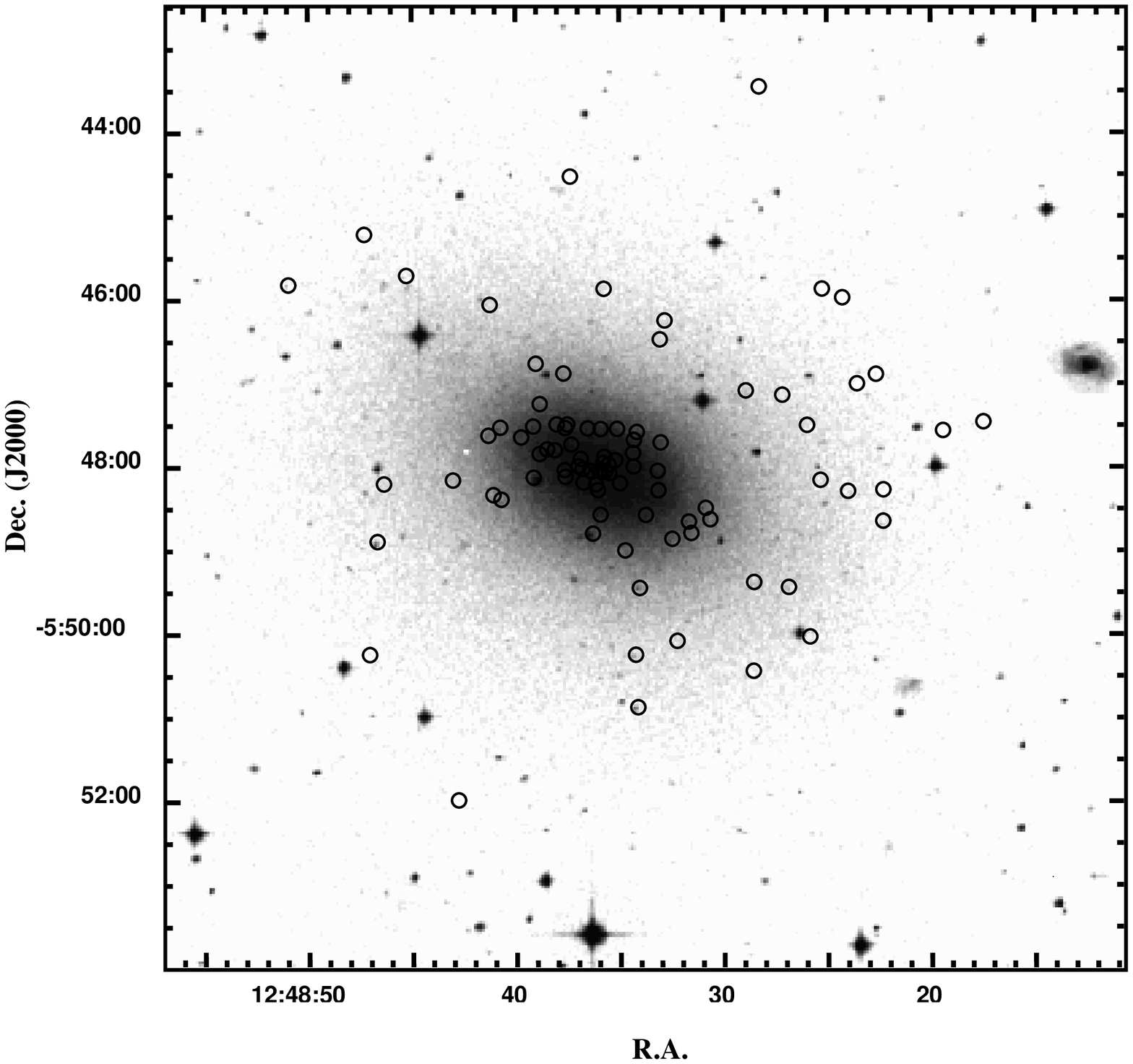}
\figcaption{
The Digital Sky Survey optical image of NGC~4697, showing the same field as in
Figure~\protect\ref{fig:xray_smo} or
Figure~1 in Paper I.
The circles show the positions of the discrete X-ray sources listed
in Table~\protect\ref{tab:src}.
\label{fig:opt}}

\vskip0.2truein

\section{Resolved Sources} \label{sec:src}

\subsection{Detections} \label{sec:src_detect}

The discrete X-ray source population on the ACIS S3 image
(Figure~1 of Paper I)
were determined, using using both
cell detection and wavelet detection algorithms.
We used the {\sc ciao celldetect} and {\sc wavdetect} programs.
The two techniques gave consistent source lists with nearly identical fluxes.
For sources located in the dense central region of NGC~4697 where
confusion and the effect of an enhanced background due to unresolved
sources or diffuse emission were important, the wavelet technique worked
slightly better.
The wavelet source detection threshold was set at $10^{-6}$,
which implies that $\la$1 false source
(due to a statistical fluctuation in the background) would be detected
in the entire S3 image.
This corresponded to requiring that the sources be detected at the
$\ge$3-$\sigma$ level.
The {\sc celldetect} threshold was initially taken to be $\ge$2-$\sigma$,
but the resulting source list was culled to remove sources whose
signal-to-ratio was less than 3-$\sigma$.
While variations in the background, exposure, and instrumental
Point-Spread-Function (PSF) across the image cause the minimum detectable
flux to vary somewhat with position, over most of the galaxy the
minimum detectable flux was about $2.6 \times 10^{-4}$ cts s$^{-1}$
($L_X = 5.0 \times 10^{37}$ ergs s$^{-1}$ at the NGC~4697 distance)
in the 0.3--10 keV band.
Fluxes were corrected for exposure and the instrument PSF.
The detection limit is higher at the edges of the field, and at the
southern and western ends of the image, where the instrumental PSF is
larger than near the aim point.
All of the source detections were verified by examining the image.

%
% Table src
%
%
% table of X-ray sources
%
\begin{table*}[p]
\tiny
\begin{center}
\caption{Discrete X-ray Sources \label{tab:src}}
\begin{tabular}{lcccccccccl}
\tableline
\tableline
Src.&Name& R.A.& Dec.& $d$ & $a$ & Count Rate& SNR & $L_X$ (0.3-10 keV)& ISB&\cr
No..&&(h:m:s)& ($\arcdeg$:$\arcmin$:$\arcsec$)& (arcsec) & (arcsec) &
($10^{-4}$ s$^{-1}$)& & ($10^{37}$ ergs s$^{-1}$) & Src. & Notes \cr
(1)&(2)&(3)&(4)&(5)&(6)&(7)&(8)&(9)&(10)&(11)\cr
\tableline
 1&CXOU J124835.8-054802&12:48:35.87&$-$5:48:02.4&\phn\phn0.00&\phn\phn0.00&   \phn41.88$\pm$3.35&   12.51&   \phn80.35& 7&a,b,c\cr
 2&CXOU J124836.0-054803&12:48:36.06&$-$5:48:03.0&\phn\phn2.80&\phn\phn3.64&\phn\phn5.03$\pm$1.17&\phn4.32&\phn\phn9.66&  &b,g\cr
 3&CXOU J124835.6-054803&12:48:35.68&$-$5:48:03.7&\phn\phn3.13&\phn\phn3.12&   \phn13.88$\pm$1.96&\phn7.09&   \phn26.63&  &b,c\cr
 4&CXOU J124836.1-054801&12:48:36.17&$-$5:48:01.9&\phn\phn4.51&\phn\phn4.87&\phn\phn5.03$\pm$1.19&\phn4.21&\phn\phn9.65&  &b\cr
 5&CXOU J124835.7-054757&12:48:35.72&$-$5:47:57.6&\phn\phn5.32&\phn\phn9.16&\phn\phn3.60$\pm$1.11&\phn3.25&\phn\phn6.91&  &b\cr
 6&CXOU J124835.9-054756&12:48:35.98&$-$5:47:56.3&\phn\phn6.26&\phn\phn9.32&   \phn18.39$\pm$2.23&\phn8.23&   \phn35.28&8?&b\cr
 7&CXOU J124836.5-054802&12:48:36.54&$-$5:48:02.6&\phn\phn9.90&   \phn11.46&\phn\phn3.19$\pm$0.99&\phn3.23&\phn\phn6.11&  &b\cr
 8&CXOU J124835.9-054752&12:48:35.93&$-$5:47:52.1&   \phn10.36&   \phn16.53&   \phn16.78$\pm$2.13&\phn7.88&   \phn32.18&8?&b\cr
 9&CXOU J124835.4-054754&12:48:35.40&$-$5:47:54.5&   \phn10.65&   \phn17.69&\phn\phn9.20$\pm$1.63&\phn5.65&   \phn17.66&  &b,c,g\cr
10&CXOU J124836.6-054801&12:48:36.60&$-$5:48:01.9&   \phn10.89&   \phn11.81&\phn\phn4.41$\pm$1.11&\phn3.97&\phn\phn8.45&  &b\cr
11&CXOU J124836.3-054811&12:48:36.32&$-$5:48:11.7&   \phn11.51&   \phn19.54&\phn\phn3.99$\pm$1.08&\phn3.69&\phn\phn7.66&  &b\cr
12&CXOU J124835.1-054811&12:48:35.16&$-$5:48:11.1&   \phn13.71&   \phn14.72&\phn\phn5.77$\pm$1.29&\phn4.46&   \phn11.07&  &b\cr
13&CXOU J124836.2-054815&12:48:36.23&$-$5:48:15.9&   \phn14.50&   \phn25.02&\phn\phn3.80$\pm$1.05&\phn3.61&\phn\phn7.29&  &b\cr
14&CXOU J124836.9-054800&12:48:36.97&$-$5:48:00.8&   \phn16.36&   \phn17.77&   \phn16.93$\pm$2.16&\phn7.85&   \phn32.47&  &b,c\cr
15&CXOU J124836.9-054810&12:48:36.93&$-$5:48:10.5&   \phn17.69&   \phn26.02&\phn\phn3.33$\pm$1.02&\phn3.37&\phn\phn6.38&  &b\cr
16&CXOU J124837.1-054758&12:48:37.17&$-$5:47:58.8&   \phn19.68&   \phn20.61&   \phn11.32$\pm$1.76&\phn6.43&   \phn21.71&  &h\cr
17&CXOU J124837.0-054753&12:48:37.05&$-$5:47:53.4&   \phn19.74&   \phn19.84&\phn\phn4.61$\pm$1.16&\phn3.96&\phn\phn8.85&  &b\cr
18&CXOU J124834.5-054759&12:48:34.51&$-$5:47:59.0&   \phn20.62&   \phn25.80&\phn\phn3.16$\pm$1.02&\phn3.10&\phn\phn6.07&  &b,g\cr
19&CXOU J124834.5-054749&12:48:34.54&$-$5:47:49.3&   \phn23.87&   \phn36.72&   \phn15.62$\pm$2.06&\phn7.60&   \phn29.97&  &g,h\cr
20&CXOU J124837.7-054806&12:48:37.79&$-$5:48:06.3&   \phn28.91&   \phn35.65&\phn\phn9.33$\pm$1.61&\phn5.81&   \phn17.90&  &\cr
21&CXOU J124837.8-054801&12:48:37.83&$-$5:48:01.6&   \phn29.25&   \phn32.93&\phn\phn3.28$\pm$1.02&\phn3.22&\phn\phn6.28&  &\cr
22&CXOU J124834.5-054739&12:48:34.50&$-$5:47:39.9&   \phn30.42&   \phn50.44&   \phn19.32$\pm$2.27&\phn8.53&   \phn37.07&  &\cr
23&CXOU J124836.0-054732&12:48:36.09&$-$5:47:32.1&   \phn30.47&   \phn48.20&\phn\phn4.89$\pm$0.71&\phn4.10&\phn\phn9.37&  &\cr
24&CXOU J124837.5-054743&12:48:37.51&$-$5:47:43.2&   \phn31.07&   \phn33.09&   \phn15.92$\pm$2.06&\phn7.72&   \phn30.55&  &\cr
25&CXOU J124836.1-054833&12:48:36.11&$-$5:48:33.5&   \phn31.35&   \phn52.58&\phn\phn6.65$\pm$1.49&\phn4.47&   \phn12.76&  &\cr
26&CXOU J124835.3-054732&12:48:35.30&$-$5:47:32.0&   \phn31.55&   \phn54.08&\phn\phn8.36$\pm$1.51&\phn5.55&   \phn16.04&  &g\cr
27&CXOU J124836.7-054731&12:48:36.72&$-$5:47:31.6&   \phn33.22&   \phn46.68&   \phn31.25$\pm$2.86&   10.91&   \phn59.95&  &g\cr
28&CXOU J124834.3-054734&12:48:34.35&$-$5:47:34.2&   \phn36.15&   \phn60.75&\phn\phn8.55$\pm$1.53&\phn5.59&   \phn16.41&  &\cr
29&CXOU J124833.3-054802&12:48:33.35&$-$5:48:02.2&   \phn37.65&   \phn43.09&\phn\phn5.80$\pm$1.36&\phn4.26&   \phn11.13&  &\cr
30&CXOU J124838.3-054747&12:48:38.30&$-$5:47:47.2&   \phn39.29&   \phn39.29&\phn\phn5.80$\pm$1.29&\phn4.48&   \phn11.13&  &\cr
31&CXOU J124833.3-054816&12:48:33.32&$-$5:48:16.1&   \phn40.55&   \phn40.68&\phn\phn6.38$\pm$1.46&\phn4.37&   \phn12.23&  &\cr
32&CXOU J124837.8-054731&12:48:37.80&$-$5:47:31.3&   \phn42.32&   \phn48.83&   \phn13.02$\pm$1.87&\phn6.98&   \phn24.97&  &\cr
33&CXOU J124833.9-054833&12:48:33.93&$-$5:48:33.8&   \phn42.80&   \phn49.38&\phn\phn3.02$\pm$0.93&\phn3.25&\phn\phn5.79&  &\cr
34&CXOU J124837.7-054728&12:48:37.71&$-$5:47:28.8&   \phn43.33&   \phn51.86&\phn\phn2.89$\pm$0.92&\phn3.16&\phn\phn5.55&  &\cr
35&CXOU J124838.6-054746&12:48:38.68&$-$5:47:46.6&   \phn44.70&   \phn44.80&\phn\phn8.02$\pm$1.48&\phn5.42&   \phn15.38&  &\cr
36&CXOU J124833.2-054741&12:48:33.21&$-$5:47:41.7&   \phn44.82&   \phn66.03&   \phn51.05$\pm$3.65&   14.00&   \phn97.92& 5&g\cr
37&CXOU J124836.4-054847&12:48:36.48&$-$5:48:47.1&   \phn45.63&   \phn77.63&\phn\phn8.29$\pm$1.60&\phn5.18&   \phn15.90&  &e\cr
38&CXOU J124838.2-054728&12:48:38.22&$-$5:47:28.8&   \phn48.51&   \phn54.25&\phn\phn3.66$\pm$1.05&\phn3.50&\phn\phn7.02&  &\cr
39&CXOU J124839.0-054749&12:48:39.03&$-$5:47:49.9&   \phn48.77&   \phn49.79&\phn\phn5.69$\pm$1.27&\phn4.49&   \phn10.92&  &\cr
40&CXOU J124839.3-054806&12:48:39.32&$-$5:48:06.9&   \phn51.68&   \phn61.94&   \phn48.02$\pm$3.53&   13.61&   \phn92.11& 9&e\cr
41&CXOU J124834.9-054859&12:48:34.92&$-$5:48:59.2&   \phn58.57&   \phn88.01&   \phn12.17$\pm$1.82&\phn6.67&   \phn23.35&  &\cr
42&CXOU J124839.3-054730&12:48:39.35&$-$5:47:30.2&   \phn61.00&   \phn62.44&   \phn16.75$\pm$2.11&\phn7.95&   \phn32.13&  &\cr
43&CXOU J124839.9-054738&12:48:39.92&$-$5:47:38.0&   \phn65.18&   \phn65.19&\phn\phn7.94$\pm$1.46&\phn5.45&   \phn15.24&  &\cr
44&CXOU J124839.0-054714&12:48:39.03&$-$5:47:14.1&   \phn67.41&   \phn76.70&\phn\phn9.53$\pm$1.60&\phn5.94&   \phn18.27&  &\cr
45&CXOU J124832.6-054851&12:48:32.66&$-$5:48:51.0&   \phn68.32&   \phn77.43&   \phn11.32$\pm$1.75&\phn6.46&   \phn21.72& 4&\cr
46&CXOU J124831.8-054838&12:48:31.85&$-$5:48:38.6&   \phn70.08&   \phn71.43&\phn\phn2.76$\pm$0.90&\phn3.08&\phn\phn5.30&  &\cr
47&CXOU J124831.7-054846&12:48:31.74&$-$5:48:46.8&   \phn76.02&   \phn79.56&   \phn10.51$\pm$1.70&\phn6.19&   \phn20.15&  &\cr
48&CXOU J124837.8-054652&12:48:37.88&$-$5:46:52.2&   \phn76.35&      106.53&\phn\phn3.45$\pm$0.98&\phn3.51&\phn\phn6.62&  &\cr
49&CXOU J124831.0-054828&12:48:31.04&$-$5:48:28.8&   \phn76.82&   \phn76.99&\phn\phn4.18$\pm$1.10&\phn3.81&\phn\phn8.01&  &e\cr
50&CXOU J124840.8-054822&12:48:40.88&$-$5:48:22.7&   \phn77.41&      102.56&\phn\phn2.76$\pm$0.88&\phn3.13&\phn\phn5.30&  &\cr
51&CXOU J124840.9-054731&12:48:40.93&$-$5:47:31.0&   \phn81.74&   \phn81.77&\phn\phn5.22$\pm$1.19&\phn4.39&   \phn10.01&  &e,g\cr
52&CXOU J124841.2-054819&12:48:41.27&$-$5:48:19.4&   \phn82.28&      105.63&   \phn11.24$\pm$1.73&\phn6.50&   \phn21.56&  &g,h\cr
53&CXOU J124830.8-054836&12:48:30.82&$-$5:48:36.9&   \phn82.91&   \phn82.96&   \phn25.75$\pm$2.62&\phn9.84&   \phn49.40& 2&e,f\cr
54&CXOU J124834.2-054926&12:48:34.23&$-$5:49:26.2&   \phn87.34&      128.94&\phn\phn3.61$\pm$1.10&\phn3.28&\phn\phn6.92&  &e,f\cr
55&CXOU J124841.5-054736&12:48:41.50&$-$5:47:36.8&   \phn87.82&   \phn88.77&\phn\phn2.70$\pm$0.88&\phn3.07&\phn\phn5.17&  &e,f,g\cr
56&CXOU J124839.2-054645&12:48:39.22&$-$5:46:45.2&   \phn91.93&      117.05&\phn\phn3.93$\pm$1.07&\phn3.66&\phn\phn7.54&  &\cr
57&CXOU J124833.2-054627&12:48:33.23&$-$5:46:27.8&      102.45&      176.64&   \phn10.16$\pm$1.64&\phn6.21&   \phn19.49&  &\cr
58&CXOU J124843.2-054808&12:48:43.22&$-$5:48:08.8&      109.76&      129.36&\phn\phn4.21$\pm$1.08&\phn3.89&\phn\phn8.08&  &g\cr
59&CXOU J124833.0-054614&12:48:33.00&$-$5:46:14.3&      116.30&      200.48&\phn\phn3.87$\pm$1.02&\phn3.78&\phn\phn7.42&  &\cr
60&CXOU J124829.1-054704&12:48:29.10&$-$5:47:04.6&      116.47&      174.71&\phn\phn2.62$\pm$0.86&\phn3.06&\phn\phn5.02&  &\cr
61&CXOU J124835.9-054551&12:48:35.93&$-$5:45:51.5&      130.93&      213.59&   \phn13.68$\pm$1.89&\phn7.24&   \phn26.23&  &e,f,g\cr
62&CXOU J124832.4-055004&12:48:32.42&$-$5:50:04.2&      132.27&      184.92&\phn\phn8.56$\pm$1.64&\phn5.22&   \phn16.42&  &\cr
63&CXOU J124828.7-054922&12:48:28.72&$-$5:49:22.1&      133.23&      140.44&\phn\phn7.46$\pm$1.51&\phn4.94&   \phn14.31&  &\cr
64&CXOU J124834.4-055014&12:48:34.42&$-$5:50:14.1&      133.48&      207.13&   \phn17.34$\pm$2.24&\phn7.73&   \phn33.25&  &e,f,g\cr
65&CXOU J124827.3-054707&12:48:27.34&$-$5:47:07.7&      138.58&      197.35&   \phn26.98$\pm$2.68&   10.08&   \phn51.76& 1&\cr
66&CXOU J124841.4-054602&12:48:41.43&$-$5:46:02.9&      145.44&      181.78&\phn\phn7.86$\pm$1.44&\phn5.47&   \phn15.07&  &\cr
67&CXOU J124826.1-054729&12:48:26.15&$-$5:47:29.4&      148.80&      192.57&\phn\phn3.67$\pm$1.10&\phn3.34&\phn\phn7.05&  &e,f\cr
68&CXOU J124825.5-054808&12:48:25.51&$-$5:48:08.8&      154.85&      172.71&\phn\phn3.27$\pm$1.02&\phn3.22&\phn\phn6.28&  &\cr
69&CXOU J124827.0-054925&12:48:27.04&$-$5:49:25.7&      155.87&      159.82&   \phn23.51$\pm$2.59&\phn9.08&   \phn45.11&  &e,f\cr
70&CXOU J124846.5-054811&12:48:46.54&$-$5:48:11.5&      159.40&      187.77&\phn\phn3.58$\pm$0.99&\phn3.59&\phn\phn6.86&  &\cr
71&CXOU J124834.3-055051&12:48:34.32&$-$5:50:51.8&      170.98&      267.92&   \phn18.07$\pm$2.34&\phn7.72&   \phn34.67& 6&e\cr
72&CXOU J124846.8-054852&12:48:46.86&$-$5:48:52.9&      171.50&      231.36&   \phn82.08$\pm$4.65&   17.66&      157.46&11&e,g,i\cr
73&CXOU J124824.1-054816&12:48:24.18&$-$5:48:16.8&      175.09&      191.32&\phn\phn6.04$\pm$1.37&\phn4.40&   \phn11.60&  &\cr
74&CXOU J124828.7-055025&12:48:28.75&$-$5:50:25.8&      178.53&      219.11&\phn\phn7.24$\pm$1.50&\phn4.81&   \phn13.88&  &\cr
75&CXOU J124826.0-055001&12:48:26.02&$-$5:50:01.2&      189.03&      202.62&\phn\phn4.94$\pm$1.23&\phn4.02&\phn\phn9.48&  &e\cr
76&CXOU J124823.7-054659&12:48:23.74&$-$5:46:59.6&      191.63&      263.26&   \phn14.70$\pm$2.03&\phn7.25&   \phn28.20&  &\cr
77&CXOU J124845.4-054541&12:48:45.44&$-$5:45:41.9&      200.30&      225.34&   \phn28.64$\pm$2.72&   10.53&   \phn54.94&10&g\cr
78&CXOU J124822.4-054815&12:48:22.48&$-$5:48:15.7&      200.30&      220.53&   \phn23.52$\pm$2.60&\phn9.06&   \phn45.12&  &g\cr
79&CXOU J124822.4-054838&12:48:22.49&$-$5:48:38.2&      202.89&      212.50&\phn\phn4.51$\pm$1.25&\phn3.60&\phn\phn8.65&  &\cr
80&CXOU J124825.4-054551&12:48:25.42&$-$5:45:51.6&      203.59&      326.05&\phn\phn3.22$\pm$0.95&\phn3.41&\phn\phn6.19&  &\cr
81&CXOU J124822.8-054652&12:48:22.83&$-$5:46:52.8&      206.77&      285.40&\phn\phn6.27$\pm$1.38&\phn4.56&   \phn12.04&  &\cr
82&CXOU J124824.4-054557&12:48:24.44&$-$5:45:57.9&      211.22&      330.79&      133.53$\pm$5.98&   22.34&      256.15&  &g\cr
83&CXOU J124837.5-054430&12:48:37.54&$-$5:44:30.9&      212.96&      335.92&\phn\phn3.16$\pm$0.93&\phn3.41&\phn\phn6.06&  &\cr
84&CXOU J124847.2-055014&12:48:47.23&$-$5:50:14.0&      214.60&      339.59&\phn\phn7.29$\pm$1.46&\phn4.91&   \phn13.98&  &g\cr
85&CXOU J124847.4-054512&12:48:47.47&$-$5:45:12.4&      242.55&      272.78&   \phn16.19$\pm$3.22&\phn5.02&   \phn31.05&12&d\cr
86&CXOU J124819.6-054733&12:48:19.60&$-$5:47:33.3&      244.61&      298.39&\phn\phn3.80$\pm$1.11&\phn3.43&\phn\phn7.30&  &\cr
87&CXOU J124842.9-055158&12:48:42.97&$-$5:51:58.4&      258.63&      445.89&\phn\phn4.36$\pm$1.42&\phn3.06&\phn\phn8.36&  &\cr
88&CXOU J124851.1-054548&12:48:51.12&$-$5:45:48.6&      263.99&      268.36&\phn\phn5.15$\pm$1.22&\phn4.22&\phn\phn9.88&  &e\cr
89&CXOU J124817.6-054727&12:48:17.66&$-$5:47:27.0&      274.17&      336.35&   \phn10.22$\pm$1.80&\phn4.68&   \phn19.61&  &\cr
90&CXOU J124828.4-054326&12:48:28.43&$-$5:43:26.4&      297.46&      512.79&\phn\phn8.26$\pm$1.55&\phn5.33&   \phn15.84&  &\cr
\tableline
\end{tabular}
\end{center}
\tablenotetext{a}{\scriptsize The position of Src.~1 agrees with the optical center
of NGC~4697 to within the errors.}
\tablenotetext{b}{\scriptsize Positions and fluxes of sources near the center of NGC~4697
are uncertain due to crowding.}
\tablenotetext{c}{\scriptsize This source appears to be extended, although this may be
due to confusion with other sources or the effect of diffuse emission.}
\tablenotetext{d}{\scriptsize This source is at the edge of the S3 detector, and its
flux is uncertain due to a large exposure correction.}
\tablenotetext{e}{\scriptsize Possible faint optical counterpart.}
\tablenotetext{f}{\scriptsize Globular cluster is possible optical counterpart.}
\tablenotetext{g}{\scriptsize Source may be variable.}
\tablenotetext{h}{\scriptsize Supersoft source.}
\tablenotetext{i}{\scriptsize The optical ID of this source is an AGN at redshift
$z = 0.696$.}
\end{table*}

Table~\ref{tab:src} lists the 90 discrete sources detected by this
technique,
sorted in order of increasing distance $d$ from the center of NGC~4697.
Columns 1-9 give
the source number,
the IAU name,
the source position (R.A. and Dec., J2000),
the projected distance $d$ from the center of NGC~4697,
the projected isophotal semi-major axis $a$ (see \S~\ref{sec:src_distr} below),
the count rate and the 1-$\sigma$ error,
the signal-to-noise ratio SNR for the count rate,
and
the unabsorbed 0.3--10 keV X-ray luminosity $L_X$ (assuming the source
is located at the distance of NGC~4697, see \S~\ref{sec:src_lum} below).
The conversion to luminosities assumes the best-fit source spectrum
(thermal bremsstrahlung with $kT = 8.1$ keV and Galactic absorption with
$N_H = 2.12 \times 10^{20}$ cm$^{-2}$;
\S~\ref{sec:spectra} and Table~\ref{tab:spectra}, row~3).
The statistical errors in the positions of most of the sources are quite
small ($\sim 0.2\arcsec$),
and the overall absolute errors are probably $\sim$0.5\arcsec\ near the
center of the field, with larger errors near the outside of the field.
The values of the distance from the galaxy center $d$ are actually
computed from the position of Src.~1;
its position agrees with the position of the optical center of the
NGC~4697 to within the errors, but relative X-ray positions are more
accurate.
Thus, $d$ might differ from the distance from the optical center of
the galaxy by $\sim$1\arcsec.

As noted above, our detection limit for sources should result in $\la$1
false source (due to a statistical fluctuation) in the entire S3 field
of view.
However, many of the detected sources may be unrelated foreground or
(more likely) background objects.
Based on the source counts in
Brandt et al.\ (2000) and
Mushotzky et al.\ (2000),
we would expect about 10-15 serendipitous sources in our observation.
These should be spread out fairly uniformly over the S3 image
(Figure~1 of Paper I),
except for the effects of the reduced
exposure and increased PSF at the outside of the field.
Thus, the unrelated sources should mainly be found at larger distances
from the optical center of NGC~4697 (the bottom part of Table~\ref{tab:src}),
while the sources associated with NGC~4697 should be concentrated to the
center of the galaxy.
Within 2\arcmin\ of the center of NGC 4697 (roughly the region covered
by Figure~2 in Paper I), $\sim$2 of the $\sim$60 the detected
sources would be expected to be unrelated to NGC~4697.

\subsection{Identifications} \label{sec:src_id}

The position of Src.~1 agrees with the optical position of the
center of NGC~4697
(R.A.=12$^{\rm h}$48$^{\rm m}$35\fs71, Dec.=-5\arcdeg48\arcmin02\farcs9;
Wegner et al.\ [1996])
to within the combined X-ray and optical errors.
Its luminosity is $L_X = 8 \times 10^{38}$ ergs s$^{-1}$.
This source appears to be broader than the instrumental PSF
(Fig.~\ref{fig:xray_cen}).
On the other hand, the density of sources is quite high
near the center of NGC~4697, and the apparent extension may be due
to confusion with other sources or diffuse emission.
It is possible that the central source is due in part to an
AGN and/or to one or more LMXBs,
as discussed below in
\S~\ref{sec:discuss_agn}.

Near the center of the galaxy, the density of sources is quite high,
and it is likely that some of the positions, fluxes, and sizes are
affected by source confusion.
Src.~3 is very close to Src.~1, and also appears somewhat extended.
It may be a composite or may be affected by the central Src.~1.
Src.~2 is very close to both Src.~1 and Src.~4.
Srcs.~7 and 10 are quite close together.

ISB detected 12 X-ray sources around NGC~4697
using the {\it ROSAT} HRI, all of which lie within the field of view of
the S3 image.
All but one of these sources lie within a few arcsec of strong sources
which we detect with {\it Chandra}.
The identifications of these sources (the sources numbers in
Table~1 from ISB) are listed in column 10 of Table~\ref{tab:src}.
ISB Src.~7 corresponds to the central X-ray source (Src.~1 in the present
paper).
ISB Src.~8 was an extended source located about 7\arcsec\ north of the
center of the galaxy;
we believe this source is actually due to the superposition of our
sources 6 and 8, with possible contributions from sources 5 and/or 9 as
well.
In general, the ISB sources have {\it Chandra} S3 count rates
$\ga 2 \times 10^{-3}$ cts s$^{-1}$, and are thus almost an order
of magnitude brighter than the faintest sources we detect.
ISB Src.~3 was located about 54\arcsec\ west of the center of
NGC~4697.
We don't detect a source or any enhancement in the background surface
brightness at this location.
Given that ISB Src.~3 was the least significant source detected
by ISB (only slightly better than 2--$\sigma$), it
seems likely that this source was a statistical fluctuation
in the {\it ROSAT} HRI image.
Alternatively, this source may be variable, and might have declined by 
more than an order of magnitude since the {\it ROSAT} observation
(\S~\ref{sec:src_var}).

The positions of the X-ray sources were examined on the Digital Sky
Survey (DSS) image of this region
(Fig.~\ref{fig:opt}).
Eleven of the sources had possible faint optical counterparts on
this image;
four others appear on deeper optical images discussed below.
These are all marked with note ``e'' in Table~\ref{tab:src}.
The possible faint optical counterparts of Srcs.~71 and 72 were noted
previously by ISB.
None of the sources corresponded to objects list on NED or SIMBAD.

However, several of the X-ray sources coincide with candidate globular
clusters associated with NGC~4697.
Src.~64 has a faint optical counterpart on the
DSS image which agrees with the position of globular cluster candidate 33
from the catalog of Hanes (1977).
(However, this source has an unusual X-ray spectrum, which may indicate
that is is actually a background AGN; see \S~\ref{sec:src_colors} below.)
Most of the Hanes (1977) globulars are fairly bright and located at
relatively large distances from NGC~4697.
J. Kavelaars (2000, private communication)
gives a list of generally fainter candidate globular clusters located
between 1\farcm5 and 2\farcm5 from the center of NGC~4697.
By comparing this list with the X-ray source positions we find
faint candidate globular clusters located within 1\arcsec\  of the positions
of X-ray Srcs.~53, 54, 55, 61, 64, 67, and 69.
The globular cluster associated with Src.~64 is the same one listed
by Hanes (1977).
Given the density of candidate globulars and the very good position
agreement, one would expect $\sim$0.3 associations to occur at
random.
Thus, it is likely that all of the 7 associations of X-ray sources
with optical globular cluster candidates in Table~\ref{tab:src} are
real.
However,
at the distance to NGC~4697, globular clusters are not resolved in
ground-based optical images, and the candidate globulars were identified by
luminosities and (possibly) colors.
As a result, as many as half of them might be unrelated faint
optical objects, rather than globular clusters.

%
% table of source hardnesses
%
\begin{table*}[t]
\tiny
\begin{center}
\caption{Hardness Ratios for Sources \label{tab:colors}}
\begin{tabular}{lcc|lcc|lcc}
\tableline
\tableline
&&&&&&&&\\
Src.&H21&H31&Src.&H21&H31&Src.&H21&H31 \\
&&&&&&&&\\
\tableline
&&&&&&&&\\
 1&$-$0.22($-$0.31,$-$0.13)&$-$0.34($-$0.42,$-$0.25)&32&+0.02($-$0.17,+0.21)&$-$0.33($-$0.52,$-$0.15)&62&+0.00($-$0.30,+0.30)&$-$0.21($-$0.51,+0.10)\\
 2&+0.01($-$0.32,+0.34)&$-$0.68($-$1.00,$-$0.36)&33&$-$0.30($-$0.70,+0.10)&$-$0.71($-$1.00,$-$0.22)&63&$-$0.38($-$0.65,$-$0.12)&$-$0.18($-$0.46,+0.10)\\
 3&$-$0.44($-$0.58,$-$0.30)&$-$0.69($-$0.83,$-$0.56)&34&+0.19($-$0.31,+0.69)&$-$0.03($-$0.73,+0.67)&64&$-$0.05($-$0.20,+0.10)&$-$0.92($-$1.00,$-$0.78)\\
 4&$-$0.03($-$0.33,+0.27)&$-$0.42($-$0.71,$-$0.14)&35&$-$0.18($-$0.39,+0.04)&$-$0.53($-$0.75,$-$0.31)&65&+0.24(+0.14,+0.35)&$-$0.30($-$0.44,$-$0.15)\\
 5&$-$0.21($-$0.55,+0.12)&$-$0.09($-$0.44,+0.26)&36&+0.03($-$0.06,+0.12)&$-$0.26($-$0.35,$-$0.18)&66&+0.07($-$0.17,+0.31)&$-$0.83($-$1.00,$-$0.55)\\
 6&+0.00($-$0.16,+0.16)&$-$0.32($-$0.47,$-$0.18)&37&$-$0.16($-$0.37,+0.06)&$-$0.67($-$0.90,$-$0.45)&67&+0.10($-$0.87,+1.00)&+0.39($-$0.23,+1.00)\\
7 \& 10\tablenotemark{a}
&$-$0.11($-$0.33,+0.12)&$-$0.67($-$0.90,$-$0.44)&38&$-$0.21($-$0.55,+0.12)&$-$0.35($-$0.70,+0.00)&68&+0.30($-$0.12,+0.71)&$-$1.00($-$1.00,$-$0.07)\\
 8&+0.17(+0.01,+0.32)&$-$0.01($-$0.21,+0.19)&39&+0.19($-$0.07,+0.45)&$-$0.25($-$0.60,+0.10)&69&+0.07($-$0.07,+0.21)&$-$0.22($-$0.37,$-$0.06)\\
 9&$-$0.62($-$0.74,$-$0.49)&$-$0.87($-$1.00,$-$0.75)&40&+0.02($-$0.07,+0.12)&$-$0.23($-$0.32,$-$0.14)&70&$-$0.10($-$0.56,+0.36)&$-$0.04($-$0.53,+0.45)\\
11&$-$0.07($-$0.40,+0.27)&$-$0.34($-$0.68,+0.00)&41&$-$0.14($-$0.32,+0.04)&$-$0.47($-$0.65,$-$0.29)&71&$-$0.23($-$0.36,$-$0.10)&$-$0.67($-$0.81,$-$0.54)\\
12&$-$0.17($-$0.42,+0.07)&$-$0.45($-$0.70,$-$0.20)&42&+0.00($-$0.18,+0.18)&$-$0.21($-$0.38,$-$0.04)&72&+0.07(+0.01,+0.14)&$-$0.31($-$0.37,$-$0.24)\\
13&$-$0.27($-$0.57,+0.04)&$-$0.98($-$1.00,$-$0.59)&43&$-$0.23($-$0.47,+0.01)&$-$0.25($-$0.50,+0.00)&73&$-$0.38($-$0.65,$-$0.12)&$-$0.69($-$0.99,$-$0.38)\\
14&$-$0.08($-$0.23,+0.06)&$-$0.49($-$0.62,$-$0.35)&44&$-$0.35($-$0.55,$-$0.15)&$-$0.22($-$0.43,$-$0.01)&74&$-$0.27($-$0.46,$-$0.07)&$-$1.00($-$1.00,$-$0.80)\\
15&$-$0.50($-$0.86,$-$0.14)&$-$0.29($-$0.68,+0.09)&45&$-$0.12($-$0.33,+0.10)&$-$0.08($-$0.31,+0.14)&75&$-$0.37($-$0.59,$-$0.16)&$-$0.66($-$0.90,$-$0.43)\\
16&$-$1.00($-$1.00,$-$0.94)&$-$1.00($-$1.00,$-$0.93)&46&+0.00($-$0.87,+0.87)&+0.24($-$0.36,+0.85)&76&$-$0.04($-$0.23,+0.16)&$-$0.19($-$0.38,+0.00)\\
17&+0.00($-$0.32,+0.32)&$-$0.22($-$0.54,+0.10)&47&$-$0.06($-$0.26,+0.15)&$-$0.14($-$0.35,+0.06)&77&$-$0.42($-$0.51,$-$0.34)&$-$0.63($-$0.72,$-$0.55)\\
18&$-$0.46($-$0.78,$-$0.14)&$-$0.25($-$0.60,+0.09)&48&+0.10($-$0.37,+0.57)&$-$0.05($-$0.62,+0.52)&78&$-$0.06($-$0.20,+0.07)&$-$0.55($-$0.68,$-$0.42)\\
19&$-$0.95($-$1.00,$-$0.89)&$-$1.00($-$1.00,$-$0.96)&49&$-$0.27($-$0.57,+0.04)&$-$0.64($-$0.99,$-$0.30)&79&+0.00($-$0.53,+0.52)&+0.11($-$0.36,+0.57)\\
20&+0.36(+0.18,+0.55)&$-$0.08($-$0.40,+0.25)&50&$-$0.01($-$0.38,+0.36)&$-$0.92($-$1.00,$-$0.47)&80&+0.37($-$0.10,+0.85)&$-$0.09($-$0.99,+0.82)\\
21&$-$0.62($-$0.90,$-$0.33)&$-$0.44($-$0.74,$-$0.14)&51&+0.15($-$0.22,+0.52)&+0.13($-$0.27,+0.52)&81&+0.50(+0.21,+0.79)&+0.24($-$0.21,+0.70)\\
22&$-$0.05($-$0.21,+0.11)&$-$0.05($-$0.21,+0.12)&52&$-$1.00($-$1.00,$-$0.94)&$-$1.00($-$1.00,$-$0.92)&82&+0.82(+0.79,+0.85)&+0.83(+0.80,+0.86)\\
23&$-$0.53($-$0.80,$-$0.27)&$-$0.24($-$0.52,+0.04)&53&$-$0.10($-$0.21,+0.02)&$-$0.48($-$0.59,$-$0.37)&83&+1.00($-$1.00,+1.00)&+1.00(+0.68,+1.00)\\
24&$-$0.21($-$0.35,$-$0.07)&$-$0.48($-$0.61,$-$0.34)&54&$-$0.41($-$0.77,$-$0.05)&$-$0.42($-$0.81,$-$0.02)&84&+0.15($-$0.14,+0.45)&$-$0.05($-$0.44,+0.34)\\
25&$-$0.44($-$0.70,$-$0.17)&$-$0.31($-$0.59,$-$0.03)&55&+0.48($-$0.09,+1.00)&+0.45($-$0.16,+1.00)&85&+0.05($-$0.24,+0.33)&$-$0.98($-$1.00,$-$0.63)\\
26&$-$0.04($-$0.31,+0.22)&$-$0.28($-$0.54,$-$0.01)&56&+0.00($-$0.55,+0.54)&$-$0.05($-$0.61,+0.52)&86&$-$0.35($-$0.74,+0.04)&$-$0.69($-$1.00,$-$0.20)\\
27&+0.17(+0.07,+0.28)&$-$0.08($-$0.21,+0.06)&57&+0.03($-$0.19,+0.25)&$-$0.26($-$0.49,$-$0.03)&87&$-$0.74($-$1.00,+0.15)&+0.29($-$0.24,+0.82)\\
28&$-$0.26($-$0.45,$-$0.07)&$-$0.47($-$0.66,$-$0.27)&58&+0.35(+0.04,+0.67)&$-$0.04($-$0.61,+0.52)&88&+0.53(+0.19,+0.87)&+0.20($-$0.41,+0.80)\\
29&$-$0.28($-$0.55,$-$0.01)&$-$0.69($-$0.99,$-$0.40)&59&+0.00($-$0.42,+0.42)&$-$0.71($-$1.00,$-$0.23)&89&+0.04($-$0.20,+0.29)&$-$0.89($-$1.00,$-$0.55)\\
30&+0.00($-$0.29,+0.28)&$-$0.30($-$0.58,$-$0.03)&60&$-$0.12($-$0.66,+0.41)&$-$0.05($-$0.62,+0.52)&90&+0.89(+0.65,+1.00)&+0.86(+0.54,+1.00)\\
31&+0.13($-$0.21,+0.47)&$-$0.12($-$0.54,+0.30)&61&$-$0.44($-$0.58,$-$0.31)&$-$0.56($-$0.70,$-$0.43)&  &                  &                  \\
&&&&&&&&\\
\tableline
\end{tabular}
\end{center}
\tablenotetext{a}{Srcs.~7 \& 10 are very close together, and we report
their combined hardness ratios.}
\end{table*}

An example of such a miss-identified globular cluster is the 
the optical object associated with Src.~72.
Hanes (1977) lists this object as a candidate globular cluster (\#~24),
although it would be rather bright for a globular at the distance of
NGC~4697.
Moreover, the USNO-A2.0 optical catalog (Monet et al.\ 1998) indicates that
this source has an very blue color ($B - R \approx -0.7$), which is
bluer than even a Rayleigh-Jeans spectrum, and would be unheard of
for a globular cluster.
An optical spectrum kindly provided by
C. Mullis (2000, private communication)
shows that this object is a background AGN at a redshift of $z = 0.696$,
rather than a globular cluster.
The very blue color is due in part to a strong \ion{Mg}{2} $\lambda$2798
line in the spectrum, redshifted into the B band.

\subsection{Luminosities and Luminosity Function} \label{sec:src_lum}

The count rates for the sources were converted into unabsorbed luminosities
(0.3-10 keV) assuming that all of the sources were at the distance of NGC~4697,
which we take to be 15.9 Mpc
(Faber et al.\ [1989] and $H_0 = 50$ km s$^{-1}$ Mpc$^{-1}$).
We adopted the best-fit {\it Chandra} X-ray spectrum of the resolved
sources within the inner 1 effective radius
(Table~\ref{tab:spectra}, row 3 below).
The resulting factor for converting the count rate (0.3--10 keV)
into the unabsorbed luminosity $L_X$ (0.3--10 keV) was
$1.92 \times 10^{41}$ ergs cnt$^{-1}$.
The resulting X-ray luminosities are given in column~9 of
Table~\ref{tab:src}, and range from about $5 \times 10^{37}$ to
$2.5 \times 10^{39}$ ergs s$^{-1}$.
The cumulative luminosity function of all of the sources was shown
as a histogram in Figure~3 of Paper I.
There we found that the luminosity function of the NGC~4697
sources could not be fit by a single power law, as the
luminosity function has a ``knee''
at $L_X \approx 3 \times 10^{38}$ ergs s$^{-1}$.
A broken power-law, 
\begin{equation} \label{eq:xlum}
\frac{ d N }{ d L_{38} } = N_o \, \left\{
\begin{array}{l}
( L_{38} / L_b )^{-\alpha_l} \qquad L_{38} \le L_b \\
\\
( L_{38} / L_b )^{-\alpha_h} \qquad L_{38} >  L_b \\
\end{array}
\right. \, ,
\end{equation}
gave a good fit,
where $L_{38}$ is the X-ray luminosity (0.3--10 keV) in units
of $10^{38}$ ergs s$^{-1}$.
The best fit, determined by the maximum-likelihood method, gave
$N_o = 8.0^{+7.1}_{-5.2}$,
$\alpha_l = 1.29^{+0.36}_{-0.49}$, 
$\alpha_h = 2.76^{+1.81}_{-0.39}$,
and a break luminosity of
$L_b = 3.2^{+2.0}_{-0.8} \times 10^{38}$ ergs s$^{-1}$
We also determined the luminosity functions for sources within
the optical half-light elliptical isophote and for sources are
larger distances.
The errors were increased significantly (particularly for the outer
region), but the luminosity functions agreed with the one shown
in Figure~3 of Paper I.

\subsection{Hardness Ratios of Sources} \label{sec:src_colors}

We studied the crude spectral properties of the resolved sources by
using hardness ratios.
Hardness ratios or X-ray colors have the advantage that they can
be applied to weaker sources.
We defined two hardness ratios as H21 $\equiv ( M - S ) / ( M + S )$
and H31 $\equiv ( H - S ) / ( H + S ) $, where $S$, $M$, and $H$ are
the total counts in our soft (0.3--1 keV), medium (1--2 keV), and hard
(2--10 keV) bands, respectively.
The hardness ratios of the sources and 90\% confidence regions are listed in
Table~\ref{tab:colors} (see also Figure~4 in Paper I).
For comparison, the hardness ratios (H21,H31) are
$(-0.38,-0.57)$ for all of the emission,
$(-0.69,-0.82)$ for the unresolved emission,  and
$(-0.14,-0.37)$ for the sum of the sources, all within one effective
radius
(\S~\ref{sec:diffuse}).

There are three moderate luminosity sources with hardness ratios of
$(-1,-1)$, 
which means that they have no detectable emission beyond 1 keV.
For example, a blackbody spectrum with a temperature of 0.1 keV
and the Galactic absorption towards NGC~4697 ($N_H = 2.12 \times 10^{20}$
cm$^{-2}$) would give hardness ratios of
(H21,H31) $\approx (-0.97,-1.00)$.
These three sources,
Src.~16,
Src.~19,
and
Src.~52,
are almost certainly supersoft sources
(e.g., Kahabka \& van den Heuvel 1997).
All three are located at small enough distances from the center of
NGC~4697 that they are likely to be associated with the galaxy.
Srcs.~19 \& 52 also may be variable (\S~\ref{sec:src_var}).
We believe this observation represents the first detection of supersoft
sources in a luminous elliptical galaxy.

There are three sources with hardness ratios of $\sim$$(1,1)$,
one of which is the brightest source in the field (Src.~82),
These sources are
Src.~82,
Src.~83,
and
Src.~90.
As is clear from their source numbers, all of these sources are located
far ($>$3\farcm5) from the center of NGC~4697.
These are probably unrelated, strongly absorbed AGNs,
similar to the sources which produce the hard component of the
X-ray background, and which appear strongly at the faint fluxes in the
deep {\it Chandra} observations of blank fields
(Brandt et al.\ 2000;
Mushotzky et al.\ 2000;
Giacconi et al.\ 2001).
As a comparison, a power-law spectrum with a photon spectral index
$\Gamma = 1.5$ and with an absorbing column of $N_H = 10^{22}$ cm$^{-2}$
gives hardness ratios of
(H21,H31) $\approx (+0.88,+0.92)$.

There are eight sources (Srcs.~13, 50, 64, 66, 68, 74, 85, \& 89)
with hardness ratios near $(0,-1)$
which have essentially no hard emission.
For example, a power-law spectrum with a photon spectral index
$\Gamma = 3.5$ and with an absorbing column of
$N_H = 3 \times 10^{21}$ cm$^{-2}$
gives hardness ratios of
(H21,H31) $\approx (-0.18,-0.81)$.
Six of these sources are at large radii ($>$2\farcm2),
which suggests that most of this population is also unrelated to NGC~4697.
The hardness ratios for these sources are consistent with the
soft-band-only sources seen in the Giacconi et al.\ (2001) deep field
at similar flux levels.
However, note that Src.~64 was also identified with a candidate globular
cluster in NGC~4697;
perhaps this source is a globular cluster LMXB with an unusual X-ray
spectrum, or the optical counterpart may be a background AGN rather than
a globular.
(Note that the 11 possible background sources selected by colors in the
last two paragraphs plus the AGN [Src.~72] would account for most of the
$\sim$10--15 unrelated sources expected based on deep blank sky observations.)

Most of the sources lie in a diagonal swath centered at about
$(-0.15,-0.40)$.
Note that hardness ratios of $(-0.14,-0.37)$ correspond to the cumulative
source spectrum with one effective radius 
(Table~\ref{tab:spectra}, row 3 below).
This is a hard spectrum fit by
thermal bremsstrahlung with $kT = 8.1$ keV and Galactic absorption with
$N_H = 2.12 \times 10^{20}$ cm$^{-2}$.
These values are similar to but slightly harder than the integrated colors
or the entire galaxy, but considerably harder than the values for the
unresolved emission.
There may be some tendency for the fainter sources to have higher
$H/M$ ratios than the brighter sources.

\subsection{Variability of Sources} \label{sec:src_var}

We searched for variability in the X-ray emission of the resolved sources
over the duration of the {\it Chandra} observation using the KS test.
This test can detect a variation in the flux from the source over
the $\sim$11 hour duration of the observation, such as a secular
increase in the flux or a sudden turn-on or off of the source.
In terms of the phenomenology of LMXBs in our Galaxy, one might detect
orbital variations or other secular variations.
Although we could, in principal, detect a Type I X-ray burster,
the luminosities
(typically $\la$ the Eddington luminosity of a NS) and durations
(10-$10^3$ s) imply that we would expect $\la$ 1 count from a burst
at the distance of NGC~4697.
The KS test cannot detect short term variations in the source which do not
individually contribute significantly to the total number of counts.
For example, we could not detect periodic pulses due to rotational
modulation of an accreting neutron star.
In principal, other tests could be done to search for pulsations from
some of the brighter sources;
in practice this is difficult because of the 3.2 s periodicity on the
events imposed by the ccd readout.
We found that 11 sources
were inconsistent with a constant flux at $>$90\% confidence;
% (Srcs.~2, 9, 18, 26, 36, 51, 52, 55, 58, 64, \& 84);
these sources are noted in Table~\ref{tab:src}.
In Figure~\ref{fig:var}, we show the histogram of the total counts from
the five sources with more than 20 net counts and with variability detected
at $>$95\% confidence.

%
% Figure var
%
\centerline{\null}
\vskip3.55truein
\includegraphics{f4.eps}
\figcaption{
The solid histograms gives the accumulated fraction of events for sources as
a function of the accumulated exposure time.
The dashed line is the predicted distribution under the hypothesis that
the source plus background rate is constant.
The five sources shown are those with more than 20 net counts (subtracting
background) and with variability detected at $>$95\% confidence.
\label{fig:var}}

\vskip0.2truein

We also tested for variability over longer time scales by comparing
the Chandra count rates of the sources with detections or limits
from a long (78,744 s) {\it ROSAT} HRI observation of the galaxy on
1997 June 19 - July 20
(ISB).
We considered the sources as variable if they increased or decreased by
$>$50\% at $>$1$\sigma$ significance.
Sources with significantly higher fluxes at the time of the
{\it Chandra} observation were Srcs.~19, 27, 52, 61, 64, 72 (ISB Src.~11),
78, and 82.
Src.~64 also appeared to vary during the {\it Chandra} observation.
Both Srcs.~19 and 52 are supersoft sources.
The third supersoft source (Src.~16) also showed some evidence for
variability, but not at as significant a level.

\subsection{Spatial Distribution of Sources} \label{sec:src_distr}

One might expect the stellar X-ray sources associated with NGC~4697 to
have a spatial distribution which is very similar to that of the
optical light.
Unfortunately, the spatial distribution of detected X-ray sources
may also be affected by variations in the sensitivity limit due
to variations in exposure, the instrumental PSF, and the diffuse background
and emission.
Because of these possible problems and statistical limits due to the
number of sources, we limit this discussion to a few simple comparisons
of the X-ray source and optical distributions.
We consider only sources within 3\farcm5 of the center of NGC~4697, which
is slightly less than the distance to the nearest chip edge.

Figure~\ref{fig:pa} shows the distribution of the absolute values of the
position angle ($PA$) of the sources.
Here, $PA$ is measured from north to east, and is taken to be in the
range $-180^\circ < PA \le 180^\circ$.
For a set of sources whose projected density is constant on concentric
aligned ellipses of constant ellipticity $e$, the distribution of
angles $\phi$ relative to the semimajor axis is
\begin{equation} \label{eq:pa}
n( \phi ) \, d \phi = 
\frac{N ( 1 - e )}{2 \pi} \,
\frac{d \phi}{ (1 - e )^2 \cos^2 \phi + \sin^2 \phi} \, ,
\end{equation}
where $n( \phi ) \, d \phi$ is the number of sources with angles between
$\phi$ and $\phi + d \phi$, and $N$ is the total number of sources.
The short-dash curve in Figure~\ref{fig:pa} is the predicted distribution
based on the optical photometry of the galaxy,
plus the expected number of background sources in this region.
The optical photometry gives
$PA = 67^{\circ} \pm 4^{\circ}$ and $e = 0.42 \pm 0.06$
(Jedrzejewski et al.\ 1987;
Faber et al.\ 1989;
Peletier et al.\ 1990);
the errors approximately cover the range of values in the literature
plus the radial variation at radii greater than 10\arcsec.
The long-dash curve is the best-fit elliptical distribution plus
background, determined by a maximum-likelihood fit to the observed
values.
The best fit distribution gives
$PA = 79^\circ \pm 15^\circ$ and $e = 0.40 \pm 0.09$.
The best fit curve is not a statistically significant improvement on the
optical model.

%
%	Figure pa
%
\centerline{\null}
\vskip2.35truein
\includegraphics{f5.eps}
\figcaption{
The solid histogram gives the distribution of the position angles of
the X-ray sources within 3\farcm5 of the center of NGC~4697 in
$10^\circ$ bins, as a function of the absolute value of the position
angle $PA$.
$PA$ is measured from north to east, and is taken to be in the
range $-180^\circ < PA \le 180^\circ$.
The short-dash curve is the predicted distribution
based on the optical photometry of the galaxy,
plus the expected number of background sources in this region.
The long-dash curve is the best-fit elliptical distribution plus
background, determined by a maximum-likelihood fit to the observed
values.
\label{fig:pa}}

\vskip0.2truein

We also compared the radial distribution of sources with the photometry
of the optical light.
Figure~\ref{fig:r} shows the accumulated source number as function
of the radius.
The dashed curve is the predicted distribution if the source
counts follow the optical light in the galaxy plus the expected
number of background sources.
The optical photometry was modeled as a de Vaucouleurs profile
with effective semimajor and semiminor axes of
$a_{\rm eff} = 95\arcsec$ and $b_{\rm eff} = 55\arcsec$.
The KS test indicates that the optical distribution is a good fit
to the X-ray source distribution.
We determined the best-fit de Vaucouleurs' profile plus background
fit to the source radial distribution, but is was not distinguishable
from the optical distribution.
We also did this analysis using the elliptical semimajor axes of the
sources ($a$ values in Table~\ref{tab:src}), but the results were nearly
identical.
These comparisons show that the X-ray source distribution is elongated
in about the same direction and by about the same amount as the optical
light, and that the radial distribution of X-ray
sources is proportional to that of the optical light, all to
within the errors.

%
%	Figure r
%
\centerline{\null}
\vskip2.35truein
\includegraphics{f6.eps}
\figcaption{
The solid histogram gives the accumulated number of
X-ray sources within 3\farcm5 of the center of NGC~4697
as a function of the radius.
The dash curve is the predicted distribution
based on the optical photometry of the galaxy,
plus the expected number of background sources in this region.
\label{fig:r}}

\vskip0.2truein

\section{Diffuse Emission} \label{sec:diffuse}

\subsection{Resolved vs.\ Diffuse Emission} \label{sec:diffuse_RvsD}

We determined the portion of the X-ray emission due to resolved
sources and unresolved emission, both for the entire {\it Chandra }
X-ray band (0.3--10 keV) and for three narrower bands:
hard H (2-10 keV), medium M (1-2 keV),
and soft S (0.3-1 keV).
We found that these bands gave reasonable
count rates and spectral discrimination for a spectrum like that of the
entire galaxy (\S~\ref{sec:spectra}), which has very soft and very hard
components.
The emission was determined for two spatial regions.
First, it was done within the elliptical optical isophote
which contains
one half of the optical light from the galaxy.
We refer to these counts as coming from within ``one effective radius.''
This isophote has an effective semimajor axis of $95^{\prime\prime}$,
an effective semiminor axis of $55^{\prime\prime}$,
an ellipticity of 0.42,
and a position angle of 67$^\circ$.
We also determined the fluxes within an elliptical annular region ranging
from one to two effective radii.
The population of resolved sources was determined as discussed
above (\S~\ref{sec:src}).
Then, the total emission was determined, and the source flux subtracted
to give the amount of unresolved emission.

%
%	Table counts
%
\begin{table*}[t]
\small
\caption{Resolved vs.\ Diffuse Emission \label{tab:counts}}
\begin{center}
\begin{tabular}{llccccc}
\tableline
\tableline
&&&&Hardness&Fraction&Luminosity\cr
Region&Origin&Band&Counts&(H21 or H31)&(\%)&($10^{40}$ ergs s$^{-1}$)\cr
\tableline
$<$ 1 $a_{\rm eff}$&All       &Total& 3715$\pm$90  &                       &(100)      &1.57 \cr
               &          &  H  &  584$\pm$69  &$-0.57^{+0.04}_{-0.04}$&(100)      &\cr
               &          &  M  &  975$\pm$36  &$-0.38^{+0.03}_{-0.03}$&(100)      &\cr
               &          &  S  & 2154$\pm$107 &                       &(100)      &\cr
               &Resolved  &Total& 2271$\pm$57  &                       & 61$\pm$10 & 1.11 \cr
               &          &  H  &  474$\pm$31  &$-0.37^{+0.03}_{-0.03}$& 81$\pm$14 &\cr
               &          &  M  &  771$\pm$35  &$-0.14^{+0.03}_{-0.03}$& 79$\pm$5  &\cr
               &          &  S  & 1026$\pm$41  &                       & 48$\pm$3  &\cr
               &Unresolved&Total& 1444$\pm$107 &                       & 39$\pm$10 & 0.46 \cr
               &          &  H  &  110$\pm$76  &$-0.82^{+0.15}_{-0.09}$& 19$\pm$14 &\cr
               &          &  M  &  204$\pm$50  &$-0.69^{+0.08}_{-0.06}$& 21$\pm$5  &\cr
               &          &  S  & 1128$\pm$115 &                       & 52$\pm$3  &\cr
               &Diffuse   &Total&  869$\pm$138 &                       & 23$\pm$4  & 0.18 \cr
               &          &  S  &  869$\pm$138 &                       & 40$\pm$7  &      \cr
               &Discrete  &Total& 2846$\pm$166 &                       & 77$\pm$4  & 1.39 \cr
               &          &  H  &  584$\pm$69  & ($-0.37$)             & (100)     &\cr
               &          &  M  &  975$\pm$36  & ($-0.14$)             & (100)     &\cr
               &          &  S  & 1285$\pm$85  &                       & 60$\pm$7  &\cr
$1 - 2 a_{\rm eff}$&All       &Total& 1831$\pm$107 &                       &(100)      & 0.65 \cr
               &          &  H  &   92$\pm$223 &$-0.88^{+0.71}_{-0.11}$&(100)      &\cr
               &          &  M  &  247$\pm$33  &$-0.72^{+0.06}_{-0.05}$&(100)      &\cr
               &          &  S  & 1492$\pm$269 &                       &(100)      &\cr
               &Resolved  &Total&  365$\pm$25  &                       & 19$\pm$2  & 0.18 \cr
               &          &  H  &   68$\pm$14  &$-0.44^{+0.10}_{-0.09}$& $>$48     &\cr
               &          &  M  &  125$\pm$16  &$-0.16^{+0.08}_{-0.08}$& 51$\pm$9  &\cr
               &          &  S  &  173$\pm$18  &                       & 12$\pm$2  &\cr
               &Unresolved&Total& 1466$\pm$110 &                       & 81$\pm$2  & 0.47 \cr
               &          &  H  &   24$\pm$223 &$-0.96^{+1.95}_{-0.04}$& $<$52     &\cr
               &          &  M  &  122$\pm$37  &$-0.83^{+0.07}_{-0.05}$& 49$\pm$9  &\cr
               &          &  S  & 1319$\pm$270 &                       & 88$\pm$2  &\cr
               &Diffuse   &Total& 1110$\pm$163 &                       & 61$\pm$10 & 0.30 \cr
               &          &  S  & 1110$\pm$163 &                       & 74$\pm$17 &      \cr
               &Discrete  &Total&  721$\pm$127 &                       & 39$\pm$10 & 0.35 \cr
               &          &  H  &   92$\pm$223 & ($-0.88$)             &(100)      &\cr
               &          &  M  &  247$\pm$33  & ($-0.72$)             &(100)      &\cr
               &          &  S  &  382$\pm$68  &                       & 26$\pm$17      &\cr
\tableline
\end{tabular}
\end{center}
\end{table*}

The results are listed in Table~\ref{tab:counts}.
For each spatial region, the emission is divided into ``Resolved'' and
``Unresolved'' components, with ``All'' denoting the sum of these two.
The truly ``Diffuse'' emission is the unresolved emission, corrected
for unresolved binary sources (\S~\ref{sec:diffuse_unresolved}),
while ``Discrete'' emission is the sum of resolved and unresolved
binary sources.
The emission is divided into the soft (S), medium (M), and hard (H)
bands, with ``Total'' denoting the sum of these (all emission from
0.3 to 10 keV).
In addition to the total number of counts, we give the hardness
ratios H21 and H31 as defined in \S~\ref{sec:src_colors}.
The fractional contribution of resolved, unresolved, and truly diffuse
emission to the counts within each band is given.
Finally, the unabsorbed luminosity for each component and band are given,
based on the best-fit spectra for each component
(\S~\ref{sec:spectra} and Table~\ref{tab:spectra} below).

Within one effective radius, 61\% of the counts and 71\% of the X-ray
luminosity is resolved into individual X-ray sources for the total band.
In the hard, medium, and soft bands, the resolved count fractions are
81\%, 79\%, and 48\%, respectively.
Between 1 and 2 effective radii, the resolved fraction drops but the
spectrum of the unresolved emission gets harder, suggesting that this
is partly the result of lower sensitivity to point sources at radii where the
point spread function (PSF) is larger.
However, it is also clear that there is a very spatially extended soft 
component to the X-ray emission.

\subsection{Unresolved Binary Sources and Truly Diffuse Emission}
\label{sec:diffuse_unresolved}

A portion of the unresolved emission must also come from LMXBs which
fall below our threshold for reliable detection of resolved sources.
Indeed, if one sets the detection threshold for sources lower, one finds
many more 2-$\sigma$ fluctuations than expected just from Poisson
statistics.
However, it would be difficult to use such a fluctuation analysis to
extend the statistical detection of LMXBs down a factor of $\sim$50 to
$L_X \sim 10^{36}$ ergs s$^{-1}$, the approximate lower limit
for LMXBs in globular clusters (Hertz \& Grindlay 1983).
If the observed luminosity function of the resolved sources in
NGC~4697 (Fig.~4, Paper I) is extended down to
$L_X = 10^{36}$ ergs s$^{-1}$,
the contribution of LMXBs to the total band X-ray emission from NGC~4697
increases by about 17\%.
This result agrees approximately with the result from the X-ray colors
of the sources and unresolved emission given in the next paragraph.
Still, this is a rather significant extrapolation, which could be
inaccurate if the slope of the luminosity function changes below
$L_X = 5 \times 10^{37}$ ergs s$^{-1}$, as is seen in the bulge of M31
(Shirey et al.\ 2001).

We have estimated the portion of the unresolved emission which is due
to unresolved LMXBs based on the X-ray hardness ratios of the sources.
The spectral analysis (\S~\ref{sec:spectra}) indicates that LMXBs
produce the majority of the X-ray luminosity at photon energies above 1 keV.
Thus, we assume that all of the hard (H) and medium (M) counts in
the unresolved emission come from unresolved discrete sources.
Moreover, we assume that these unresolved sources have the same
spectral properties as the resolved sources.
We use the hardness ratios of the resolved sources and the H+M counts
of the unresolved emission to correct the soft (S) band unresolved
emission for unresolved LMXBs.
The counts and fractions for this truly ``Diffuse'' component are
listed in Table~\ref{tab:counts}.
When corrected in this way, it appears that LMXBs (resolved and unresolved)
provide 77\% of the total band counts and 89\% of the luminosity in the
inner effective radius.

The X-ray spectrum of the unresolved emission consists of two components:
a hard component with a spectrum which is consistent with that of the
resolved sources, and a soft component
(\S~\ref{sec:spec_diffuse} below).
If we attribute the hard component to unresolved sources, then we can
determine the total luminosity of the discrete sources from the luminosity
of the hard component in the total spectrum
(\S~\ref{sec:spectra} below).
Based on the spectral fits,
we would conclude that 74\% of the total luminosity is due to discrete sources.
The differences in the results using colors or spectral fits partly
reflects uncertainties in the spectral fits and count rates,
and differences in the spectral models of the different components.
However, this difference may also be due to errors
in the calibration of the ACIS S3 below 0.7 keV
(\S~\ref{sec:spectra} below).
As a result of concerns about the low energy spectral calibration, we
use the hardness ratios and counts rather than spectra to determine
the portion of the unresolved emission which is due to LMXBs.
Using the spectra requires extrapolating the fit for the spectral
range 0.7--10 keV down to 0.3 keV, where the spectral calibration
appears to be very uncertain.
Using the hardness ratios involves
only assuming that the unresolved LMXBs have the same spectrum as
resolved sources.

The remaining $\sim$11\% of the luminosity and $\sim$23\% of the
counts would come from a more diffuse component with a soft
($\sim$0.3 keV) spectrum.
In \S~\ref{sec:discuss_gas}, we show that this emission is almost certainly
due to diffuse gas.

\subsection{Spatial Distribution of Diffuse Emission}
\label{sec:diffuse_spatial}

We determined the radial distribution of the diffuse emission.
Because we are mainly interested in the (presumably gaseous) soft component,
we did this in the soft band (0.3--1.0 keV).
Resolved sources were excluded.
The unresolved emission was collected in circular annuli.
(We also determined the surface brightness in elliptical annuli
whose orientation and ellipticity matched that of the optical isophotes of
the galaxy,
and the results were essentially identical to those for circular annuli.)
The outer boundary of the largest annulus was the largest circle which
fit entirely on the S3 chip;
thus, regions of the chip corners, south edge, and west edge were not
used to determine the surface brightness.
The profile was corrected for background and for exposure.
As noted previously, the background at large radii is quite uncertain
due to the foreground emission by the North Polar Spur (\S~\ref{sec:obs}),
and we include a large systematic error in the background because of this.
As discussed in \S~\ref{sec:obs}, this systematic error covered the
range of NPS contributions from zero to all of the emission at large radii
on the S3 chip.
Thus, the errors in the resulting surface brightness values and model
fit parameters should include the full range of possible background
contributions.

%
%	Figure xsurf
%
\centerline{\null}
\vskip2.35truein
\includegraphics{f7.eps}
\figcaption{
The profile of the unresolved, soft band (0.3--1 keV) emission
as a function of the projected radius $r$.
The short dashed line shows a de Vaucouleurs profile
which fits the optical surface brightness of the galaxy.
The long dashed curve is a beta model fit, while the solid curve is
the best-fit assuming that the emission is the sum of emission
which is proportional to the optical light plus a beta model.
\label{fig:xsurf}}

\vskip0.2truein

The observed profile is shown in Figure~\ref{fig:xsurf} as a function
of the projected radius $r$.
The observed soft X-ray profile is much broader than the optical profile
for the galaxy.
The short dashed line shows a de Vaucouleurs profile
with an effective radius of $r_{\rm eff} = 72\arcsec$,
which is the azimuthally-averaged value for the optical light.
A fit to the profile with the X-ray surface brightness proportional
to the optical surface brightness is completely unacceptable.
We tried to fit the profile using the standard beta model
\begin{equation} \label{eq:beta}
I_X ( a ) = I_o
\left[ 1 + \left( \frac{r}{r_c} \right)^2 \right]^{-3 \beta + 1/2} \, ,
\end{equation}
where $r_c$ is the core radius.
This provided a acceptable fit, which is shown as the long dashed curve
in Figure~\ref{fig:xsurf}.
The value of $\beta = 0.335 \pm 0.004$ implies a rather flat profile,
although fairly flat profiles are
found in many X-ray bright elliptical galaxies
(e.g., Forman et al.\ 1985;
Trinchieri, Fabbiano, \& Canizares 1986).
On the other hand, the core radius is required to be very small,
$r_c = 2\farcs9 \pm 0\farcs2$.
This implies that the X-ray surface brightness is nearly a power-law
function of the radius, $I_X \propto r^{-1.01}$.

%
% table of X-ray spectra
%
\begin{table*}[t]
\tiny
\caption{X-ray Spectral Fits \label{tab:spectra}}
\begin{center}
\begin{tabular}{llcclccclcccrc}
\tableline
\tableline
&&&&&&&&&&&&&\\
&&&&
\multicolumn{3}{c}{Hard Component}&&
\multicolumn{4}{c}{Soft Component}&&\\
&&&&&&&&&&&&&\\
\cline{5-7} \cline{9-12}
&&&&&&&&&&&&&\\
&&&
$N_H$&
Model&
$kT_h$ or $\Gamma$&
% $F^h_X$ (0.3-10 keV)&&
$F^h_X$&&
Model&
$kT_s$ or $kT_{\rm BB}$&
Abund.&
% $F^s_X$ (0.3-10 keV)&&\\
$F^s_X$ &&\\
&&&&&&&&&&&&&\\
Row&
Origin&
Region&
($10^{20}$ cm$^{-2}$)&&
(keV)&
% ($10^{-13}$ ergs cm$^{-2}$ s$^{-1})$&&&
\tablenotemark{a}&&&
(keV)&&
% ($10^{-13}$ ergs cm$^{-2}$ s$^{-1})$&
\tablenotemark{a}&
Counts&
$\chi^2$/dof\\
&&&&&&&&&&&&&\\
\tableline
&&&&&&&&&&&&&\\
1  &Sources     &$<$ 1 $a_{\rm eff}$&(2.12)&bremss&(5.2)&$3.11\pm0.17$&&MEKAL&(0.26)&(0.07)&$<$0.11&1595&60.1/70\\
2  &Sources     &$<$ 1 $a_{\rm eff}$&(2.12)&bremss&(5.2)&$3.11\pm0.14$&&&&&(0.0)&1595&60.1/71\\
3* &Sources     &$<$ 1 $a_{\rm eff}$&(2.12)&bremss&$8.1^{+2.8}_{-1.9}$&$3.50\pm0.15$&&&&&(0.0)&1595&51.4/70\\
4  &Sources     &$<$ 1 $a_{\rm eff}$&$<$3.90&bremss&$9.1^{+3.5}_{-2.5}$&$3.49\pm0.15$&&&&&(0.0)&1595&50.3/69\\
5  &Sources     &$<$ 1 $a_{\rm eff}$&(2.12) &power &$1.57\pm0.08$&$3.96\pm0.23$&&&&&(0.0)&1595&52.9/70\\
6  &Sources     &$<$ 1 $a_{\rm eff}$&(2.12) &bremss&$9.3^{+5.0}_{-2.5}$&$3.53\pm0.19$&&bbody&$<$1.70&&$<$0.42&1595&48.8/68\\
&&&&&&&&&&&&&\\
7  &Sources     &$>$ 1 $a_{\rm eff}$&(2.12)&bremss&(8.1)&$3.39\pm0.07$&&&&&(0.0)&1586&99.8/70\\
8  &Sources     &$>$ 1 $a_{\rm eff}$&(2.12)&bremss&$30^{+39}_{-13}$&$4.42\pm0.69$&&&&&(0.0)&1586&74.5/69\\
9* &Sources     &$>$ 1 $a_{\rm eff}$&$14.2\pm5.6$&bremss&$9.7^{+6.7}_{-3.0}$&$4.38\pm0.33$&&&&&(0.0)&1586&61.8/68\\
&&&&&&&&&&&&&\\
10 &Sources     &$L_X < L_b$    &(2.12)      &bremss&$6.5^{+3.3}_{-1.8}$&$1.89\pm0.12$&&&&&(0.0)&908&37.0/44\\
11*&Sources     &$L_X < L_b$    &(2.12)      &bremss&$10.8^{+29.7}_{-4.7}$&$1.93\pm0.19$&&bbody&$0.14^{+0.10}_{-0.04}$&&$0.21^{+0.48}_{-0.21}$&908&30.8/42\\
&&&&&&&&&&&&&\\
12 &Sources     &$L_X > L_b$    &(2.12)      &bremss&$9.1^{+4.9}_{-2.6}$&$2.03\pm0.12$&&&&&(0.0)&906&40.2/40\\
13*&Sources     &$L_X > L_b$    &$9.0^{+7.9}_{-7.1}$&bremss&$6.4^{+4.2}_{-2.1}$&$2.03\pm0.30$&&&&&(0.0)&906&38.3/39\\
&&&&&&&&&&&&&\\
14 &Unresolved  &$<$ 1 $a_{\rm eff}$&(2.12)&bremss&(5.20)&$0.39^{+0.07}_{-0.23}$&&MEKAL&(0.26)&(0.07)&$1.19\pm0.18$&705&79.9/105\\
15 &Unresolved  &$<$ 1 $a_{\rm eff}$&$<$22.83&bremss&(5.20)&$0.29^{+0.14}_{-0.17}$&&MEKAL&(0.26)&(0.07)&$1.52^{+1.18}_{-0.36}$&705&78.5/104\\
16 &Unresolved  &$<$ 1 $a_{\rm eff}$&(2.12)&bremss&$1.5^{+24.9}_{-1.0}$&$0.29^{+0.99}_{-0.22}$&&MEKAL&$0.25^{+0.11}_{-0.03}$&$>$0.03&$1.12^{+0.22}_{-0.99}$&705
&77.5/102\\
17*&Unresolved  &$<$ 1 $a_{\rm eff}$&(2.12)&bremss&(8.1)&$0.27^{+0.22}_{-0.21}$&&MEKAL&$0.29^{+0.10}_{-0.07}$&$0.06\pm0.04$&$1.16^{+0.08}_{-0.13}$&705&
79.4/103\\
&&&&&&&&&&&&&\\
18*&Total       &$<$ 1 $a_{\rm eff}$&(2.12)&bremss&(5.20)&(3.45)&&MEKAL&(0.26)&(0.07)&(1.22)&2372&135.0/156\\
19  &Total       &$<$ 1 $a_{\rm eff}$&$<$3.07&bremss&(5.20)&(3.45)&&MEKAL&(0.26)&(0.07)&(1.22)&2372&134.4/155\\
20 &Total       &$<$ 1 $a_{\rm eff}$&(2.12)&bremss&$5.7^{+2.6}_{-1.5}$&$3.62\pm0.26$&&MEKAL&$0.22^{+0.09}_{-0.04}$&$>$0.02&$1.46\pm0.31$&2372&131.4/151\\
&&&&&&&&&&&&&\\
21 &Total       &1--2 $a_{\rm eff}$&(2.12)&bremss&(8.1)&$0.38^{+1.09}_{-0.21}$&&MEKAL&$0.26^{+0.13}_{-0.06}$&(0.07)&$1.78^{+0.49}_{-1.01}$&817&193.7/243\\
&&&&&&&&&&&&&\\
\tableline
\end{tabular}
\end{center}
\tablenotetext{a}{Units are $10^{-13}$ ergs cm$^{-2}$ s$^{-1}$ in 0.3--10
keV band.}
\tablenotetext{*}{The adopted best-fit model for this emission.}
\end{table*}

The spectrum and colors of the unresolved emission indicate that
a portion of this emission is due to unresolved stellar sources.
The very small value of the core radius in the beta model fit
also suggests that the emission near the center may have a significant
contribution from unresolved stellar sources with a cuspy de Vaucouleurs
profile.
Thus, we also tried fitting the surface brightness profile with the sum
of a de Vaucouleurs profile and beta model, with the effective radius
of the de Vaucouleurs profile fixed by the optical profile as before.
This led to a fit which was only marginally better than the single
beta model
($\Delta \chi^2 = -1.10$ for one additional fitting parameters),
shown as the solid curve in
Figure~\ref{fig:xsurf}.
The normalizations of the two components are roughly consistent with
the decomposition of the unresolved emission into truly diffuse and
discrete emission in Table~\ref{tab:counts} in the various regions
based on X-ray colors.
In this fit, the values of $r_c$ and $\beta$ are both very large and
poorly constrained,
$r_c > 296\arcsec$ and
$\beta > 1.24$.
Obviously, these values are poor determined since the core radius is
larger than the largest radius at which the surface brightness is
determined.
This suggests a rather flat surface brightness distribution for the
truly diffuse emission out to a large radius.
On the other hand, a constant surface brightness is not an acceptable
fit for the truly diffuse emission
($\Delta \chi^2 = +12.85$ for two fewer fitting parameters).
The surface brightness due to diffuse emission is fit acceptably with
a fairly flat power-law distribution with 
$I_X \propto r^{-0.55}$.

In conclusion, the surface brightness distribution of the diffuse emission
is uncertain due to corrections for the unresolved point sources and the
background including the NPS.
However, it has a radially declining surface brightness which is
considerably flatter than that of the optical stellar emission from the
galaxy.
Note that this is true even if the diffuse emission is not corrected
for the unresolved point sources
(Fig.~\ref{fig:xsurf}), and that the very extended emission occurs at
surface brightness levels which exceed those of the background and the
NPS by approximately one order of magnitude.

\section{Spectral Analysis} \label{sec:spectra}

At the time when this analysis was done, there were considerable
uncertainties in the soft X-ray spectral response of the S3 chip below
0.7 keV
(e.g., Markevitch et al.\ 2000).
We found significant residual deviations in most of the spectral
fits for any reasonable spectrum at energies $\la$0.7 keV.
Also, we have previous determined the spectrum of NGC~4697 with
both the {\it ROSAT} PSPC and {\it ASCA}
(ISB).
We found that the {\it Chandra} spectrum at low energies was
inconsistent with the {\it ROSAT} PSPC or the joint {\it ROSAT}
and {\it ASCA} spectra.
In addition to the spectra presented in ISB, we extracted the total
{\it ROSAT} PSPC spectrum for the inner effective radius
for comparison to the equivalent {\it Chandra} S3 spectrum for exactly
the same region
(\S~\ref{sec:spec_sources}).
(We could not determine the {\it ASCA} spectrum of the same
region as it is smaller than the spatial resolution of {\it ASCA}.)
The {\it Chandra} and {\it ROSAT} spectra were completely inconsistent
at soft X-ray energies;
they agree very well above 0.71 keV, and very poorly below this
energy.
Thus, we restrict all of our spectral analysis to the energy range
0.72--10.0 keV.
In order to allow $\chi^2$ statistics to be used, all of the spectra were
grouped to at least 20 counts per spectral bin.

Previous {\it ROSAT} and {\it ASCA} spectra of early-type galaxies have
indicated that they have at least two spectral components,
a very hard component which may be due to X-ray binaries and/or and AGN
(Matsumoto et al.\ 1997;
Allen et al.\ 2000),
and a softer component.
In X-ray luminous early-type galaxies, the soft component is dominant,
and it is clearly due to diffuse gas at a temperature of $\sim$1 keV
(Forman et al.\ 1985;
Canizares, Fabbiano, \& Trinchieri 1987).
In X-ray faint early-type galaxies, the soft component is much softer,
and its origin is still uncertain
(Fabbiano, Kim, \& Trinchieri 1994;
Pellegrini 1994;
Kim et al.\ 1996;
Irwin \& Sarazin 1998a,b).
The {\it ASCA} spectrum of the hard component has generally been fit by
either a power-law (characterized by a photon spectral index $\Gamma$,
where $\Gamma > 0$ implies a photon spectrum which declines with energy;
Allen et al.\ 2000)
or by a thermal bremsstrahlung spectrum
(characterized by a hard component temperature $T_h$;
Matsumoto et al.\ 1997).
The soft component in X-ray bright galaxies is usually fit by
the MEKAL model for the emission from a low density, optically thin
plasma
(Irwin \& Sarazin 1998a,b)
This model is characterized by the temperature of the gas ($T_s$)
and by the abundances of the heavy elements.
Given the limited statistics we have in our spectra, we will assume
that the heavy element abundances have the solar ratios, and only allow
the overall abundance of the heavy elements to vary.
In X-ray faint galaxies, it is unclear what the appropriate soft emission
model should be.
If the soft emission is due to diffuse gas, then the MEKAL model would
again be appropriate.
If it is due to an optically thick stellar component (including the
same LMXBs which produce the hard component), then it might be
better represented as a blackbody, characterized by a temperature ($T_s$
again).
Thus, we have used a spectral models which include both a hard (power-law
or bremsstrahlung) and soft (MEKAL or blackbody) component.
In ISB, the spectrum of NGC~4697 was fit with the sum of a hard
bremsstrahlung and soft MEKAL model.

Models were fit to the spectra using XSPEC.
The results are summarized in Table~\ref{tab:spectra},
where the errors are at the 90\% confidence level.
This Table gives the absorbing column $N_H$, hard spectral model,
bremsstrahlung temperature $T_h$ or power-law photon index $\Gamma$,
unabsorbed flux of the hard component $F^h_X$ (0.3--10 keV), soft component
model, soft component temperature $T_s$, abundance (relative to solar)
for the MEKAL component, unabsorbed flux of the soft component $F^s_X$
(0.3--10 keV), number of net counts in the spectrum,
and $\chi^2$ per degree of freedom (dof).
The Galactic absorbing column towards NGC~4697 is
$N_H = 2.12 \times 10^{20}$ cm$^{-2}$
(Stark et al.\ 1992).

\subsection{X-ray Spectra of Resolved Sources} \label{sec:spec_sources}

The background for the resolved sources came from regions located around
each source with three times the area of the regions for the source
spectrum.
We first determined the X-ray spectrum of all of the resolved sources
within one effective radius.
The resulting spectrum, which gave about 1600 counts in the 0.72--10 keV energy
range after subtraction of background,
is shown in Figure~\ref{fig:spec_sources}.

We first fit the spectrum assuming the same spectra properties and
properties as were used to fit the total {\it ASCA} plus {\it ROSAT}
spectrum in ISB (row 1, Table~\ref{tab:spectra}).
The normalizations of the hard (bremss) and soft (MEKAL) components were
allowed to vary.
This gave a good fit to the spectrum of the sources, but the flux of the
soft component was very weak;
the fit only gave an upper limit which was much lower than the hard flux.
Thus, we tried removing the soft component completely in the fit (row 2).
This gave a fit which was just as good ($\Delta \chi^2 < 0.001$) with
one less free parameter.
Thus, we conclude that, on average, the resolved sources in the NGC~4697
have no soft component to their spectra, and that any soft component
in the overall galaxy spectrum must come from a distinct, unresolved
component, such as interstellar gas.

%
%	Figure spec_sources
%
\vskip2.45truein
\includegraphics{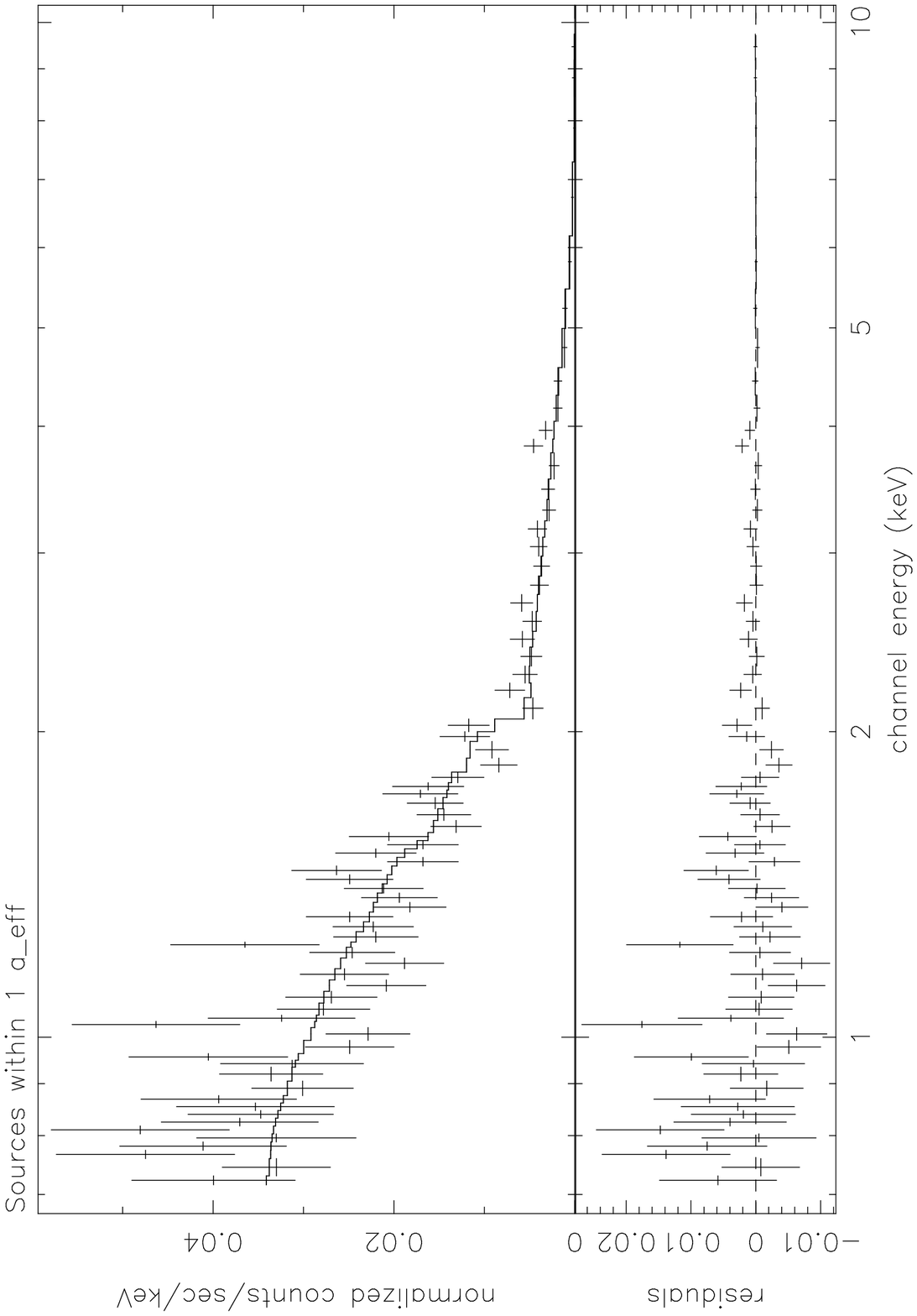}
\figcaption{
The data points in the upper panel are the cumulative X-ray spectrum of
all of the resolved sources in the inner effective radius of NGC~4697.
The solid histogram is the best-fit model spectrum
(row 3 in Table~\protect\ref{tab:spectra}).
The lower panel shows the residuals to the fit.
\label{fig:spec_sources}}

\vskip0.2truein

Next, we allowed the temperature $T_h$ of the hard component to vary
(row 3).
This resulted in a somewhat higher temperature than had been given
by the total {\it ASCA} plus {\it ROSAT} spectrum, which did not resolve the
spectra of the sources separately.
The new fit was a considerable improvement, with
$\Delta \chi^2 = 8.7$ for only one more free parameter, which is
significant at the 99.9\% level according to the f-test.
We will adopt this as our best-fit model for the resolved source
population within the inner effective radius.
This fit is shown with the data in Figure~\ref{fig:spec_sources}.
Since this region combines good statistics with a high probability
that the sources are associated with NGC~4697, we will also use this
fit as our standard model for the spectrum of the LMXB population of
NGC~4697.

Next, we allowed the absorbing column to vary to see if there was
evidence of any excess absorption beyond the Galactic column
(row 4).
We can only give an upper limit to the absorbing column, which is somewhat
greater than the Galactic value.
Presumably, this is partly due to removing the softest energies
from the spectra due to the calibration uncertainties discussed above.
Since allowing the absorbing column did not result in a greatly
improved fit
($\Delta \chi^2 = 1.1$ for one more free parameter),
we will fix the absorption at the Galactic value.

Finally, we considered different models for the hard and soft components.
We represented the hard component by a power-law rather than
by bremsstrahlung (row 5).
This resulted in a slightly worse fit than with bremsstrahlung, so we
will use the bremsstrahlung model for the hard component due to LMXBs.
Then, we tried using a blackbody model for the soft component.
This produced a better fit than the MEKAL model, but still no soft
component was required at the 90\% confidence level.

We also determined the cumulative X-ray spectrum of the resolved sources
outside of one effective radius (rows 7--9).
As was true of the inner sources, these spectra are well-fit without
any soft component in the spectrum.
The cumulative spectra of the outer sources is considerably harder than
that of the inner sources, and is fit best with an absorbing column
which is much higher than Galactic (row 9).
In row 7, we give the results of fitting the spectrum of the outer sources
using the same spectra parameters as fit the inner sources (row 3).
This did not provide a very good fit.
Rows 8 \& 9 show the result of freeing the hard component temperature
$T_h$ and the absorbing column.
The inner source spectral model can be rejected with very high confidence
($\gg$99\%).
We believe that this difference is mainly due to the greater influence of
background sources in the outer regions of the S3 chip.
Many of the background sources have hardness ratios which indicate that
they have very hard, strongly absorbed spectra
(\S~\ref{sec:src_colors}, Figure~4 in Paper I).
For example, Src.~82 is the brightest X-ray source on the S3 chip, and
provides about one quarter of the counts in the spectrum of outer sources.
It has a very hard, strongly absorbed spectrum.

We also tried to fit the spectra of the brightest and faintest sources
separately (rows 10--13).
Because the luminosity function shows a break at a characteristic
luminosity $L_b$, we separated sources into brighter sources ($L_X > L_b$)
and fainter sources ($L_X < L_b$).
In order to include as many sources as possible to improve the statistics,
but avoid too many background sources, we considered all sources within
an elliptical optical isophote with a semimajor axis of 2 $a_{\rm eff}$.
For the fainter sources, the best fit spectra were consistent with the
Galactic column, so we fixed the absorption at this level.
Unlike the case for the integrated spectra of all the sources within
one effective radius, the spectral fit was improved significantly by
including a soft component.
Either a MEKAL or blackbody spectrum did almost exactly as well.
However, the abundance in the MEKAL model was essentially unconstrained,
and there is no clear evidence for emission lines in the spectrum,
so we adopt a blackbody model for the soft component.
Rows 10 and 11 of Table~\ref{tab:spectra} show the fits without and
with the soft component.
The flux of the soft component is small, but it improves the fit
significantly.
The spectra of the brighter sources did not require any soft component.
On the other hand, they have a higher absorbing column and a lower
bremsstrahlung temperature.
The net effect is that the fainter sources may have more very soft
emission, while the brighter sources have more emission at 1-2 keV.

\subsection{X-ray Spectra of Unresolved Emission} \label{sec:spec_diffuse}

We also determined the X-ray spectra of the unresolved emission
in NGC~4697.
Because the unresolved emission has a low surface brightness, and the
background is high and uncertain due to the superposed North Polar Spur
(\S~\ref{sec:obs}), we were only able to extract a useful spectrum for the
central portion of the unresolved emission.
Thus, we will only consider the unresolved X-ray spectrum from within
the inner one effective radius;
even in this region, the net counts in the spectrum after subtraction
of background are only $\approx$700, which is marginal for spectral
analysis.
The spectrum was extracted from the inner one effective radius region,
excluding small regions around each of the sources.
Because the excluded regions also contain unresolved emission,
the fluxes we derive will be somewhat smaller than the total fluxes
for unresolved emission in the same region
(e.g., as given in Table~\ref{tab:counts}).
The background for this spectrum and the spectra of the total
emission (\S~\ref{sec:spec_total})
were taken from a combination of blank sky and outer S3 chip
spectra in order to remove the effects of the North Polar Spur,
as discussed above (\S~\ref{sec:obs}).
We include a systematic error in the background, which means that the
best-fit values of $\chi^2$ will be somewhat smaller than expected for
uncorrelated gaussian statistical errors.

%
%	Figure spec_diffuse
%
\vskip2.45truein
\includegraphics{f9.eps}
\figcaption{
The spectrum of unresolved emission from within the inner effective radius.
The notation is the same as in 
Figure~\protect\ref{fig:spec_sources}.
The solid histogram is the {\it ROSAT} adapted spectral model for the
unresolved emission
(row 17 in Table~\protect\ref{tab:spectra}).
\label{fig:spec_diffuse}}

\vskip0.2truein

We started by fitting this spectrum with the spectrum derived from
the {\it ROSAT} spectrum of the same region, which consisted of a
bremsstrahlung hard component and a MEKAL soft component (row 14).
Both the hard and soft component are required to fit the spectrum,
but the hard component is fairly weak.
The unresolved emission is much softer than the emission from the
resolved sources.
Allowing the absorbing column to vary did not improve the fit very
significantly and resulted in only an upper limit on $N_H$ (row 15).
Thus, we will keep the column fixed at the Galactic value.

If all of the spectral parameters are allowed to vary (row 16), the fit
is not significantly better, and the temperature of the hard bremsstrahlung
component is nearly unconstrained.
Under the assumption that the hard component is from fainter discrete sources
(presumably LMXBs) with a spectrum similar to that of the resolved
sources, we fix the temperature of the hard component to the value (8.1
keV) found for the resolved sources in the same region (row 3).

This fit is shown in row 17 and Figure~\ref{fig:spec_diffuse}.
The observed spectrum shows the (mainly)
\ion{Fe}{17} ($\sim$0.72 keV emitted),
\ion{Fe}{17} (0.826 keV emitted),
and the 
\ion{Ne}{10}, \ion{Fe}{17}, \ion{Fe}{21} ($\sim$1.02 keV)
line complexes,
which indicates that the soft unresolved emission is due to diffuse
gas.
At lower energies where the spectral calibration is uncertain, the
\ion{O}{8} line at 0.654 keV is clearly seen.
Some of the line ratios in the data appear to be inconsistent
with the model;
for example, the \ion{Fe}{17} (0.826 keV emitted) line complex is
too weak in the model.
This may indicate that there are calibration problems with the spectra
even above 0.72 keV, or that the model is incorrect in some way.

The fraction of the emission in the hard component is smaller than
expected from the analysis of the X-ray colors
(Table~\ref{tab:counts}) or the spatial distribution of the unresolved
emission
(\S~\ref{sec:diffuse_spatial}).
This may be the result of uncertainties in the soft X-ray spectral
response of the ACIS S3 detector, as noted above.
Also, the regions used to exclude sources from the unresolved emission
remove a higher fraction of the area near the center of the galaxy,
and there is evidence that the unresolved emission gets softer as
the radius increases.

%
%	Figure spec_total
%
\vskip2.45truein
\includegraphics{f10.eps}
\figcaption{
The total (sources plus unresolved emission) spectrum from within
the inner effective radius.
The notation is the same as in 
Figure~\protect\ref{fig:spec_sources}.
The solid histogram is the {\it ROSAT} spectral model for the
same region
(row 18 in Table~\protect\ref{tab:spectra}).
\label{fig:spec_total}}

\vskip0.2truein

\subsection{Total X-ray Spectra} \label{sec:spec_total}

We also determined the total (resolved and unresolved emission) X-ray
spectra for various regions in NGC~4697.
We first consider the total X-ray spectrum from within the inner one
effective radius.
Initially, we fit this spectrum using exactly the {\it ROSAT} spectrum (ISB)
from the same region (row 18).
This model included a bremsstrahlung hard component and a MEKAL soft
component.
As discussed above, this gave a terrible fit if the energy range used
was 0.3-10.0 keV, but gave a good fit if the soft X-ray channels were
dropped (for the range 0.72-10.0 keV).
The observed spectrum and the {\it ROSAT} model fit are shown in
Figure~\ref{fig:spec_total}.
We then tried freeing the normalizations of the hard and soft components
but leaving the shape of the two components fixed at the {\it ROSAT}
form.
This did not improve the fit, and the normalization were not changed
significantly.
We also tried allowing the absorbing column to vary (row 19);
this did not improve the fit significantly, so we kept it fixed at
the Galactic value.
We then allowed all of the spectral parameters except $N_H$ to vary (row 20);
this did not provide a significantly improved fit given the
five additional free parameters.
Moreover, the abundance of the soft MEKAL emission was very poorly determined;
since most of the soft emission in the spectrum is in lines, it is only
the product of the MEKAL normalization times the abundance which is fixed
by the spectrum.

As was the case with the unresolved spectrum, the total spectrum shows
lines due to (mainly)
\ion{O}{8} (0.654 keV emitted),
\ion{Fe}{17} ($\sim$0.72 keV emitted),
\ion{Fe}{17} (0.826 keV emitted),
and the 
\ion{Ne}{10}, \ion{Fe}{17}, \ion{Fe}{21} ($\sim$1.02 keV)
line complex.
As was true of the unresolved spectrum,
the \ion{Fe}{17} (0.826 keV emitted) line complex is too weak in the model.
Figure~\ref{fig:spec_total} also shows the Si K$\alpha$ line
at about 1.8 keV in the data but not in the model.
This is a background feature produced by fluorescence in {\it Chandra};
its presence in the spectrum after subtraction of background indicates
that the background subtraction is not perfect.
The same feature is seen in the unresolved spectrum
Figure~\ref{fig:spec_diffuse}.

At larger radii, the decreasing surface brightness of the emission,
increasing contribution of the background, and the systematic uncertainty
in the background make it more difficult to extract useful spectral
information.
We extracted the spectrum for the region from 1 to 2 effective radii
(row 21).
This spectrum had only about 800 net counts after correction for the
background.
The temperature of the hard component was not well determined, but
was consistent with that found for the sources in the inner region
(8.1 keV), so we adopt this value.
The abundance in the soft component was not well-determined
but consistent with the value from {\it ROSAT} and {\it ASCA} of
0.07, so we adopt this.
The resulting spectral fit shows that the spectrum is more dominated
by the soft emission than at smaller radii.
There is no evidence for a significant temperature gradient, although
the errors are large.

%
%	Table spectra
%

\section{Discussion} \label{sec:discuss}

\subsection{Nature of the Central Source} \label{sec:discuss_agn}

The position of Src.~1 agrees with the optical position of the
center of NGC~4697 to within the combined X-ray and optical errors.
This suggests that this source is actually an AGN.
The luminosity of this source,
$L_X = 8 \times 10^{38}$ ergs s$^{-1}$, is higher than most LMXBs, but
there are four sources which are brighter in the S3 field.
Of these, the two brightest sources in the field (Srcs.~72 \& 82)
are probably background AGNs (Src.~72 definitely is), while the other
two (Srcs.~36 \& 40) are probably LMXBs in NGC~4697.
On the other hand, the hardness ratios for the central source
(H21 $= -0.22^{+0.09}_{-0.09}$, H31 $= -0.34^{+0.09}_{-0.08}$)
are very typical of those for the LMXBs
(Table~\ref{tab:colors}, Figure~4 in Paper I), which have average hardnesses
ratios of
(H21,H31) $= (-0.14,-0.37)$ within the inner effective radius.
Thus, based on the X-ray evidence alone, the central source might
either be the active nucleus in NGC~4697, or one or more LMXBs seen
in projection against the nucleus.
The central source appears to be slightly extended, which might indicate
that it is the result of more than one point source.

NGC~4697 is not a radio source at a fairly restrictive level
(Birkinshaw \& Davies 1985).
We were unable to find any other clear evidence for nuclear activity
in the literature.
This might argue that the central source is not due to an AGN.

Stellar dynamical measurements indicate that NGC~4697 has a
central black hole with a mass
$M_{\rm BH} = 1.6 \times 10^8 \, M_\odot$, assuming our adopted distance
of 15.9 Mpc
(Gebhardt et al.\ 2000).
If we take the observed X-ray luminosity of the central source to be
an upper limit to the luminosity of an associated active nucleus, then
the AGN luminosity is $\le 4 \times 10^{-8}$ of the Eddington luminosity
of the central black hole.

\subsection{The LMXB Population} \label{sec:discuss_LMXB}

\subsubsection{X-ray to Optical Luminosity Ratio of LMXBs}
\label{sec:discuss_LMXB_x2opt}

If one compares the discrete source (resolved and unresolved LMXBs)
X-ray luminosity from within one $a_{\rm eff}$ with the optical
luminosity within the same region,
one finds an X-ray--to--optical ratio for the LMXBs of
$L_X$(LMXB, 0.3--10 keV)$/L_B = 8.1 \times 10^{29}$ ergs s$^{-1}$
$L_{B\odot}^{-1}$.
For comparison, we found a value of
$L_X$(LMXB, 0.3--10 keV)$/L_B = 7.2 \times 10^{29}$ erg s$^{-1}$
$L_{\odot}^{B}$ in the S0 galaxy NGC~1553
(Blanton, Sarazin, \& Irwin 2001),
although the systematic errors are larger in that case because the
observation only samples the top of the X-ray luminosity function.
Chandra observations of other ellipticals appear to show larger
variations in the X-ray to optical ratio
(R. Mushotzky \& L. Angelini, private communication).
For the observed luminosity function in NGC~4697 (eq.~[\ref{eq:xlum}]),
most of the luminosity comes from sources near the break luminosity.
In NGC~4697, there are only $\sim$15 of these.
Thus, part of the variation in the X-ray to optical ratios may be due
to statistical fluctuations in the number of bright LMXBs,
or to individual temporal variations in their X-ray emission.

A recent {\it XMM-Newton} observation of the bulge of M31 resolves
$\sim$90\% of the emission into LMXBs within a 5\arcmin\ radius
of the center
(Shirey et al.\ 2001).
Using our spectral model to convert to our passband, and determining
the optical luminosity in this region from the surface photometry of
Walterbos \& Kennicutt (1988), we find an
X-ray to optical luminosity ratio of
$L_X$(LMXB, 0.3--10 keV)$/L_B = 6.0 \times 10^{29}$ ergs s$^{-1}$
$L_{B\odot}^{-1}$.
Previously, Irwin \& Sarazin (1998a) used {\it ROSAT} observations
of the bulge of M31
(Primini, Forman, \& Jones 1993;
Supper et al.\ 1997)
to determine the X-ray luminosity.
Assuming 90\% of the emission is from LMXBs (based on the {\it XMM}
result) and converting to our passband gives
$L_X$(LMXB, 0.3--10 keV)$/L_B = 6.5 \times 10^{29}$ ergs s$^{-1}$
$L_{B\odot}^{-1}$,
in very good agreement with the {\it XMM} result.
However,
the X-ray to optical ratio in the bulge of M31 seems to be about
35\% smaller than that in NGC~4697.
This may be due to the steep luminosity function in M31, which causes
there to be no very bright LMXBs (\S~\ref{sec:discuss_LMXB_lumfunc}).
Sources with $L_X \ga L_b$ contribute much of the emission
in NGC~4697.

Previous {\it ROSAT} observations of the bulge of the Sa galaxy
NGC~1291 indicate an X-ray to optical ratio of
$L_X$(LMXB, 0.3--10 keV)$/L_B = 8.9 \times 10^{29}$ ergs s$^{-1}$
$L_{B\odot}^{-1}$
(Irwin \& Sarazin 1998a),
if one converts from the {\it ROSAT} hard band of 0.52--2.02 keV
using our best-fit source spectrum.
This agrees with the value for NGC~4697.
We have confirmed this determination of the X-ray--to--optical
ratio in NGC~1291 in a recent {\it Chandra} observation
(Irwin, Sarazin, \& Bregman 2001).
However, about 30\% of the source emission in NGC~1291 comes from the very
bright central point source, which is probably a central AGN.  
If this central point source is removed to determine the X-ray luminosity
due to LMXBs, then the X-ray--to--optical ratio in NGC~1291 is similar
to that in the bulge of M31, and about 50\% smaller than in NGC~4697.
It appears that there are variations in the LMXB X-ray--to--optical
ratios of early-type galaxies and spiral bulges.

\subsubsection{Luminosity Function} \label{sec:discuss_LMXB_lumfunc}

The differential luminosity function of LMXBs in NGC~4697 is well-fit by a
broken power law (eq.~[\ref{eq:xlum}])
over the range of luminosities from $5 \times 10^{37}$ to $10^{39}$ ergs
s$^{-1}$.
The break luminosity is $L_b = 3.2^{+2.0}_{-0.8} \times 10^{38}$ ergs s$^{-1}$,
which is similar to the Eddington luminosity
for spherical accretion onto a 1.4 $M_\odot$ neutron star
($2 \times 10^{38}$ ergs s$^{-1}$ for hydrogen accretion).
Of course, in addition to the statistical uncertainties, the value of
$L_b$ is affected by systematic uncertainties in the distance to NGC~4697
and in the conversion from counts to luminosity.
This agreement suggests that the sources with luminosities above the
break luminosity are accreting black holes, while those below the break
are predominantly neutron stars.
This would imply that NGC~4697 contains $\ga$15 luminous
X-ray binary systems containing black holes.
If the more luminous of these systems ($L_X \sim 10^{39}$ ergs s$^{-1}$)
are limited by the Eddington luminosity, they must contain fairly massive
($M \ga 8 \, M_\odot$) black holes.

If the luminosity function were a broken power-law which curved up
rather than down (an ``elbow'' rather than a ``knee''),
the break might be due to the simple superposition of two power-law source
distributions.
Then, the apparent break in the luminosity function would just be the
point where the two distributions crossed, and wouldn't necessarily
have any physical meaning.
However, such a superposition obviously cannot produce a broken power-law
which curves downward (a knee), as is observed in NGC~4697.
This requires that there be at least two distinct source populations above
and below the break luminosity.

We have found similar broken-power law luminosity functions in the
S0 galaxy NGC~1553
(Blanton et al.\ 2001), and the bulge of the early-type spiral NGC~1291
shows a sharp break right at the Eddington-break, with no bright non-nuclear
sources above this break
(Irwin et al.\ 2001).
Apparently, similar broken-power-laws are found to fit the LMXB populations
in X-ray bright ellipticals as well
(R. Mushotzky \& L. Angelini 2000, private communication).
In all cases, the break luminosity agrees with the Eddington luminosity
of a 1.4 $M_\odot$ neutron star, to within the errors.
Thus, this appears to be a universal feature of the LMXB populations
in optically luminous, old stellar systems.

In principle, this ``Eddington-break'' luminosity could be used as a
distance estimator for galaxies.
Margon \& Ostriker (1973) originally proposed that X-ray sources
would be limited by the Eddington luminosity and that this might be
useful as a distance estimator.
Also, the use of the peak luminosity of X-ray bursts as a distance
indicator in our Galaxy had been suggested
(e.g., van Paradijs 1978),
since Type~I burst are also thought to be Eddington-limited.
The Eddington luminosity depends on very simple physics (electron
scattering and gravity), and is determined solely by the mass of the
accreting object and the mean mass per electron.
In practice, the composition dependence resolves itself into
two interesting cases, hydrogen-dominated plasmas (the usual case)
or heavy element dominated plasmas (helium or heavier), which might occur
if the donor star had lost its outer hydrogen-rich envelope.
Although the physics of stellar core collapse is complex, the fact that
most neutron stars in binaries have masses of approximately 1.4 $M_\odot$
is presumably related to the Chandrasehkar mass, which is also determined
only by simple quantum mechanics, gravity, and slightly by composition.
Thus, the apparent universality of the break luminosity of LMXBs in early-type
galaxies is understandable in terms of very simple physics.

Unfortunately, it isn't possible to determine the break luminosity in an
individual galaxy with great statistical accuracy, because the number of
bright LMXBs is very limited (about 15 in NGC~4697).
For example, the statistical errors in the flux associated with the break
luminosity in NGC~4697 imply that the distance could only be determined to
about $\pm$20\%\ if the actual break luminosity were accurately known.
Other distance estimators in current use give statistical errors of
about 10\% for early-type galaxies
(e.g., Tonry et al.\ 2001).
One situation where the Eddington break luminosity might be particularly
useful would be in a group or cluster of galaxies, where one could determine
the total luminosity function for all of the LMXBs in the early-type
galaxies, and derive an average distance for the group or cluster.

Prior to {\it Chandra} and {\it XMM-Newton}, the Population II system
with the best-determined LMXB X-ray luminosity function was the bulge of M31
(Primini, Forman, \& Jones 1993;
Supper et al.\ 1997;
Garcia et al.\ 2000;
Shirey et al.\ 2001).
The recent {\it XMM-Newton} observation shows that
the cumulative luminosity function of bulge sources in M31 is well-fit by
a broken power-law, with an exponent of
$-0.47 \pm 0.03$ for $36.2 \le \log L_X < 37.4$ and
$-1.79 \pm 0.26$ for $37.4 \le \log L_X < 38.1$
(Shirey et al.\ 2001).
This applies to the 0.3-12 keV band, which is similar to the one
we use (0.3-10 keV).
These results are in good agreement with the luminosity function for
the bulge of M31 previously determined by {\it ROSAT}
(Primini, Forman, \& Jones 1993;
Supper et al.\ 1997).
The brightest source in the bulge of M31 has a luminosity of
$L_X \sim L_b$, so we can't compare the upper end of the luminosity
function in NGC~4697 with M31.
On the other end, our source detection limit was $\log L_X \ge 37.7$, so we
can't compare to the low luminosity end of the luminosity function of the
bulge of M31.
However, we can compare luminosity functions in the rather narrow range
$37.7 \le \log L_X \le 38.07$ (0.3-10 keV).
In this range, the exponent of the differential luminosity function
(eq.~[\ref{eq:xlum}])
in the bulge of M31 is $\alpha = 2.79 \pm 0.26$ (1-$\sigma$ errors),
while the exponent for NGC~4697 is $\alpha_l = 1.29^{+0.36}_{-0.49}$
(90\% errors).
Thus, the luminosity function in M31 declines much more steeply with
luminosity.
It seems that this rapid decline, combined with the much smaller optical
luminosity of the bulge of M31, might account for the lack of brighter
X-ray sources in M31.

\subsubsection{Globular Clusters and LMXBs} \label{sec:discuss_LMXB_globulars}

We have identified 7 X-ray sources at radii of $>$1\farcm5 in NGC~4697
with candidate globular clusters.
Because globulars are not resolved in ground based observations at the
distance of NGC~4697, some of these candidate globulars might actually
be background AGNs.
Indeed, we have found that one of the candidate globular clusters
(Hanes 1977, cluster \#~24, our Src.~72).
is a background AGN at $z = 0.696$
(C. Mullis 2000, private communication).
At present, there are no lists of globular clusters in the central regions
($\la$1\farcm5) of NGC~4697, where the diffuse optical brightness of
the galaxy makes it difficult to identify globular clusters in
ground-based observations.
Unfortunately, most of the X-ray sources (55 out of 90) are within 
1\farcm5, and within this region very few of these sources are
expected to be unrelated to NGC~4697.
A {\it Hubble} Space Telescope study to detect globular
cluster candidates in the central region of NGC~4697 would be very
useful.

Beyond 1\farcm5, roughly 20\% of the X-ray sources are associated with
globular cluster candidates.
On the other hand, the candidate globular clusters contain about
0.1\% of the optical light of the galaxy in this region.
Thus, the chance of an optical star being the donor in a LMXB is
a factor of $\sim$200 larger for stars in globular clusters than for
field stars in NGC~4697.
This indicates that globular clusters in this elliptical galaxy
are a very hospitable environment for active compact binaries, as
is true in our own Galaxy
(e.g., Hertz \& Grindlay 1983;
White, Nagase, \& Parmar 1995).
This is generally believed to result from stellar dynamical interactions
in globular clusters, which can produce compact binary systems.

One intriguing suggestion is that globular clusters might be the only
location where very compact low mass binaries can evolve, and that all
of the LMXBs in old stellar systems may have been formed in globulars
(White, Kulkarni, \& Sarazin 2001).
Of course, not all of the LMXBs are presently found in globulars, either
in the bulge of our Galaxy or in NGC~4697.
The field LMXBs might have been ejected from globular clusters by
kick velocities resulting from supernovae, by stellar dynamical processes,
or by the dissolution of the globular due to tidal effects.

Above, we have noted that there appear to be variations in
the LMXB X-ray to optical luminosity ratio of early-type galaxies.
If LMXBs all are produced in globulars, then the total luminosity of
LMXBs might correlate better with the globular cluster population
rather than the total optical luminosity.
Since the specific frequency of globular clusters (the number per
optical luminosity) varies from galaxy to galaxy, this might
explain the variation in the LMXB X-ray--to--optical ratio.
If one compares the globular cluster population in NGC~4697 with the
bulge of M31, NGC~4697 has $8^{+7}_{-4}$ times as many globulars,
and its specific globular cluster frequency $S_N$ per unit optical
luminosity is $\sim$5 times higher
(Harris 1991).
The X-ray luminosity of LMXBs is about 5 times higher in NGC~4697,
and the X-ray--to--optical ratio is about 1.35 times higher.
Thus, the LMXB luminosity ratio is intermediate between the ratio
of the optical luminosities and the ratio of the numbers of globular
clusters.
Obviously, there are too few galaxies with well-determined LMXB populations
to allow a really meaningful test at present, but there should be many more
from {\it Chandra} and {\it XMM-Newton} within the next few years.
One confusing issue is the fact that the luminosity function of LMXBs
also appears to vary widely (\S~\ref{sec:discuss_LMXB_lumfunc}), which
requires something more complex than simple scaling of the entire
population with optical luminosity or globular cluster population.

In general, the specific frequency of globular clusters $S_N$ increases
from late to early-type spirals, from spirals to S0s, from S0s to
ellipticals, and from normal giant ellipticals to cDs
(e.g., Harris 1991).
If LMXBs all are born in globular clusters, then one might expect
the specific frequency of LMXBs and their X-ray to optical ratio to
increase in the same sequence.
Already, {\it Chandra} observations may suggest an increase in the
frequency of LMXBs in going from normal ellipticals to cDs
(R. Mushotzky \& L. Angelini 2000, private communication).

\subsection{Interstellar Gas Emission} \label{sec:discuss_gas}

\subsubsection{Nature of the Diffuse Emission} \label{sec:discuss_gas_diffuse}

The unresolved X-ray emission associated with NGC~4697 appears to consist
of (at least) two components:
unresolved LMXBs and truly diffuse emission.
The X-ray colors of the unresolved emission, its X-ray spectrum,
and its spatial distribution are all consistent with this decomposition.
The colors and spatial analysis indicate that
$\approx$40\% of the counts and $\approx$61\% of the luminosity
is due to unresolved LMXBs.
A similar result is found by extrapolating the luminosity function of
the LMXBs to lower luminosities.
The X-ray spectrum gives a smaller fraction of unresolved LMXBs, but this
might be due to the uncertain soft response of the ACIS S3 or to the
varying area of the galaxy removed in order to exclude sources.

There are several arguments which indicate that the remaining truly diffuse
emission is due to hot interstellar gas.
First, this emission has a much more extended radial distribution
than the optical light in the galaxy
(Fig.~\ref{fig:xsurf}),
which suggests that it is not from stellar sources.
Second, the image of the diffuse emission
(Fig.~\ref{fig:xray_smo}, and Fig.~2 in Paper I)
shows that it is somewhat irregular, with an L-shaped morphology.
Such asymmetries are more easily produced in the distribution of gas
(e.g., by ram pressure) than in the distribution of stellar sources.
The diffuse emission is also much rounder than the highly elliptical
distribution of optical light.
Third, the diffuse emission is predominantly soft, with most of the
emission below 1 keV.
Finally, the X-ray spectrum of the unresolved emission shows low-ionization
X-ray lines from ions such at \ion{O}{8} and \ion{Fe}{17}.
This clearly implies that much of the soft emission is from an optically
thin gas.

The surface brightness distribution of this soft gaseous component
is very broad and flat, and the
total mass does not appear to converge within the S3 field of view.
As a result, it is difficult to determine the total mass of the
interstellar gas.
This problem is exacerbated by the foreground emission from the
North Polar Spur, which adds a (presumably) fairly flat emission component
at a similar temperature.
However, we have estimated the mass within a spherical region
with a 75\arcsec\ radius, which corresponds to the average effective
radius.
The mass of interstellar gas is about $1.8 \times 10^8$ $M_\odot$ within
this region.
This is much smaller than the masses of hot gas typically found
in X-ray bright ellipticals
(e.g., Forman, Jones, \& Tucker 1985).
However, the total mass is probably a factor of $\ga$3 larger given
the flat surface brightness distribution.

\subsubsection{Origin of the Soft Component in X-ray Faint Ellipticals}
\label{sec:discuss_gas_soft}

X-ray faint early-type galaxies have a very soft X-ray spectral component
whose origin has been uncertain.
One suggestion was that this emission came from the same LMXBs which
produce the hard component
(Irwin \& Sarazin 1998a,b).
Now, the LMXBs in NGC~4697 do provide a significant part of the soft
X-ray emission by simple virtue of the fact that they dominate the
total emission.
Certainly, some of the discrete X-ray sources in NGC~4697, the supersoft
sources, do have a very soft X-ray spectral component.
However, the cumulative spectrum of the sources does not show
a significant soft X-ray spectral component, and is well-fit
by a hard thermal bremsstrahlung spectrum with a reasonably high temperature
($kT_h = 8.1$ keV; \S~\ref{sec:spec_sources}).
Similar results were found for the S0 galaxy NGC~1553
(Blanton et al.\ 2001).
While there are still considerable uncertainties in the soft X-ray
spectral response of the ACIS detectors on {\it Chandra}, it appears
very unlikely that the LMXBs can account for the very soft spectral
component seen previously with {\it ROSAT} in either NGC~4697 or
NGC~1553.
By extension, it appears that the soft emission in X-ray faint ellipticals
is not primarily due to LMXBs.

The {\it XMM-Newton} spectra of the LMXBs in the bulge of M31 also
shows that, taken together, they do not have a strong soft spectral
component
(Shirey et al.\ 2001).
This confirms the {\it ROSAT} result of
Borozdin \& Priedhorsky (2000),
although Irwin \& Bregman (1999) had reached the opposite conclusion.

The soft spectral component might also be due to fainter stellar sources
such as active M stars or RS CVn binaries
(e.g., Pellegrini 1994).
There were strong energetic and spectral arguments against any of these
sources
(Pellegrini \& Fabbiano 1994; Irwin \& Sarazin 1998a).
If the soft emission were due to faint stellar sources, it would be
expected to follow the optical distribution in NGC~4697 quite closely.
In fact, the soft component is more spatially extended, rounder, and too
irregular to be due to a large number of faint stellar sources
(\S~\ref{sec:diffuse_spatial}).

The spatial distribution of the soft diffuse emission in both NGC~4697 and
NGC~1553 implies that it is due to diffuse gas.
Even more importantly, its X-ray spectrum is dominated by low ionization
X-ray lines from ions like \ion{O}{8} and \ion{Fe}{17}.
This clearly shows that the soft spectral component in these two early-type
galaxies is due to diffuse interstellar gas.
The recent {\it XMM-Newton} observation of the central bulge of M31 also
shows that there is diffuse emission with an emission-line spectrum
(Shirey et al.\ 2001).
Thus, it appears that the source of the soft component in early-type
galaxies is cool ($kT \approx 0.2-0.3$ keV) interstellar gas, as
suggested by
Pellegrini \& Fabbiano (1994).

\subsubsection{Physical State of the Hot Interstellar Gas}
\label{sec:discuss_gas_ism}

The interstellar gas in NGC~4697 is rather cool ($kT = 0.29$ keV)
in comparison to X-ray bright ellipticals.
A similar result was found for the S0 galaxy NGC~1553
(Blanton et al.\ 2001),
the bulge of the Sa galaxy NGC~1291 (Irwin et al.\ 2001),
and for the diffuse gas in the bulge of M31
(Shirey et al.\ 2001).
NGC~4697 has a low velocity dispersion for a bright elliptical,
$\sigma = 165$ km s$^{-1}$
(Faber et. al.\ 1989).
On the other hand, this galaxy also has a very high rotation velocity for
an elliptical
(Peletier et al.\ 1990),
so the velocity dispersion may not be representative of the full depth
of the galactic potential.
The observed gas temperature is consistent with the temperature vs.\ velocity
dispersion ($kT-\sigma$) correlation derived from $ROSAT$ observations of
elliptical galaxies 
(Davis \& White 1996),
although this relationship was derived including a {\it ROSAT}
single-temperature determination
for NGC~4697 which is higher than the {\it Chandra}
temperature of the gas (presumably because LMXBs and gas could not be
separated in the {\it ROSAT} spectrum).

Why are the gas temperature and gas X-ray luminosity so low in NGC~4697
(and other X-ray faint ellipticals)?
Pellegrini \& Fabbiano (1994) argue that the low temperature is a
result of a shallow gravitational potential, as indicated by the
low velocity dispersion.
As noted above, this is consistent with the observed temperature and
$\sigma$, although the strong rotation in this galaxy implies that the 
gravitational potential may be considerably deeper than expected from
the velocity dispersion.

What is the dynamical state of the interstellar gas in NGC~4697?
X-ray bright ellipticals are believed to have global cooling inflows
(e.g., Sarazin 1990).
A global inflow can occur if the radiative cooling time of the
gas is shorter than the age of the galaxy throughout the bulk of the
optical image of the galaxy.
Based on the best-fit model for the radial distribution of the soft
X-ray emission, the central gas electron number density is about
$n_e = 8.8 \times 10^{-3}$ cm$^{-3}$.
The central cooling time in the gas is about $4 \times 10^{8}$ yr, which
is much shorter than the likely age of the galaxy.

The total X-ray luminosity of the gas may be too low to be consistent
with a global cooling inflow involving the total stellar mass
loss rate of the galaxy.
The total bolometric luminosity of the gas is
$L_{\rm bol} \approx 3.6 \times 10^{40}$ ergs s$^{-1}$, assuming the
best-fit soft component model for the unresolved emission
(Table~\ref{tab:spectra}).
Given its optical luminosity and assuming a normal elliptical galaxy
stellar mass loss rate, the total rate of stellar mass loss is
about 0.5 $M_\odot$ yr$^{-1}$
(e.g., Sarazin 1990).
The heat input to the gas which is just associated with the motions
of the gas-losing stars is $3 \sigma^2 / 2$ per unit mass, which leads
to a total heating rate of
$\dot{E}_{\rm heat} \ge 1.3 \times 10^{40}$ ergs s$^{-1}$.
If the gas involved in a cooling inflow, then infall in the galactic
potential and adiabatic compression increase the heating rate by a
factor of $\sim$3.
Type Ia supernova could also increase the heating rate by at least as
large a factor.
Thus, for a global cooling inflow one would expect a bolometric
luminosity of $\ga 8 \times 10^{40}$ ergs s$^{-1}$.
At best, this would be be marginally allowed by the observed luminosity.

However, if the gas were involved in a global cooling flow, one would
expect the gas X-ray surface brightness to be much more centrally peaked
than is observed
(Fig.~\ref{fig:xsurf}).
In global cooling flow models, the X-ray surface brightness is generally
at least as centrally peaked as the optical surface brightness
(e.g., Sarazin \& White 1988;
Sarazin \& Ashe 1989).
Thus, a global cooling flow seems unlikely.
On the other hand,
if the interstellar gas were in a global supersonic outflow or wind,
the expected X-ray luminosity ($\la 10^{39}$ ergs s$^{-1}$)
would be much smaller than is observed
(D'Ercole et al.\ 1989).

Pellegrini \& Fabbiano (1994) suggest that the gas in NGC~4697 is
in a partial wind, with a cooling inflow at the center and an outflow
in the outer regions.
This could explain the low luminosity and gas mass in this galaxy.
On the other hand, in these models the X-ray surface brightness of the
gas is always more centrally condensed than the optical surface brightness
(Pellegrini \& Ciotti 1998),
which is exactly the opposite of what we observe in NGC~4697
(Fig.~\ref{fig:xsurf}).

Among simple spherical hydrodynamical models of isolated galaxies, the only
models which seem consistent with the low X-ray luminosity, low
temperature, short cooling time, and broad distribution of the gas in
NGC~4697 would seem to be subsonic inflation models prior to
cooling inflow
(Ciotti et al.\ 1991;
David, Forman, \& Jones 1991).
However, these models would probably require that much of the gas
in the galaxy have been removed $\la 10^9$ yr ago to explain the
low mass of interstellar gas today, and also why this galaxy is at
such a early hydrodynamical stage when it appears to have a very old
stellar population.

Pellegrini (1994) suggested that S0 galaxies and other rapidly rotating
early-type galaxies might be more susceptible to winds and other outflows
as a result of rotation.
She showed that there was an apparent anticorrelation between the
X-ray to optical ratio of early-type galaxies and their fractional
rotational support.
NGC~4697 might be an example of the effects of rapid rotation;
this could help to explain the low mass of ISM gas.
On the other hand, in this model one might expect the soft X-ray
image to be highly elongated in the same direction as the optical
image, which doesn't really seem to be the case
(Figure~\ref{fig:xray_smo}).

Alternatively, it may be that the gas in NGC~4697 has been strongly
affected by ram pressure or other hydrodynamical effects its environment
(e.g., White \& Sarazin 1991).
The extended morphology of the gas and its distorted morphology
would suggest motion of the galaxy to the northwest relative to the ambient
intergalactic medium.
If the motion also has a significant component along the line of sight,
ram pressure might explain both the extended distribution and the
distorted morphology.
Of course, ram pressure stripping would also explain the low X-ray
luminosity and low gas mass.

The best-fit abundance for the gas in NGC~4697 is rather low (0.06 solar),
but is very poorly determined since much of the soft X-ray emission is
due to lines.
In any case, the abundance is consistent with the abundance-temperature
correlation found by
Davis \& White (1996).

\subsubsection{Relation to Cooler Interstellar Medium}
\label{sec:discuss_gas_cooler}

NGC~4697 contains a small dust lane near its center
(Goudfrooij et al.\ 1994;
van Dokkum \& Franx 1995),
which is most prominent in images with the {\it Hubble} Space Telescope
(e.g., Dejonghe et al.\ 1996).
The dust lane is fairly regular, and appears to be part of a disk of
material which is elongated along the optical major axis of the galaxy.
There is no obvious corresponding feature in the X-ray image
(Figure~\ref{fig:xray_cen} and Paper I, Figure 2).

NGC~4697 also contains an arc-shaped region of H$\alpha$ emission, mainly
extending to the northeast of the galaxy center
(Goudfrooij et al.\ 1994).
Again, there is nothing which obviously corresponds to this in the
X-ray image
(Figure~\ref{fig:xray_cen} and Paper I, Figure 2).
There does appear to be a bit more diffuse emission and more sources
on this side of the galaxy, but this is not a very strong effect.
The spatial decomposition of the unresolved emission suggests that
it is dominated by unresolved LMXBs in the central parts of the galaxy.
Thus, there may be too little hot gas to interact with the cooler
ISM, and/or any spatial features due to interaction of diffuse X-ray
gas with cooler ISM may be washed out by emission by unresolved
LMXB emission.

If the columns of cooler gas are significant, one might expect to see
absorption features in the X-ray image corresponding to locations of
cooler ISM.
The lack of such features suggests that the columns of cooler gas are
not very large ($\la 6 \times 10^{20}$ cm$^{-2}$).

\section{Conclusions} \label{sec:conclude}

The origin of the X-ray emission in X-ray faint elliptical galaxies
has been a mystery.
Unlike X-ray bright early-type galaxies where the emission is dominated
by interstellar gas with $kT \approx 1$ keV, 
X-ray faint galaxies have both a very hard ($\sim$5 keV) and very
soft ($\sim$0.2 keV) spectral components.
The hard component was thought to be due to LMXBs or an AGN.
The origin of the very soft component was very uncertain;
LMXBs (Irwin \& Sarazin 1998a,b),
cool interstellar gas
(Pellegrini \& Fabbiano 1994),
or faint stellar sources, such as M stars and RS CVn binaries
(e.g., Pellegrini 1994) have all been suggested.

Our high spatial resolution {\it Chandra} observation of the X-ray faint
elliptical galaxy NGC~4697 resolves most of the X-ray counts
(61\% within one effective radius) into point sources.
A total of 90 individual sources are detected, of which $\sim$80
are low mass X-ray binaries (LMXBs) associated with NGC~4697.
Much of the emission is resolved even in the softest band.
The dominance of LMXBs implies that this and other X-ray faint ellipticals
have lost much of their interstellar gas.

On the other hand, NGC~4697 does have a modest amount of X-ray
emission by hot interstellar gas.
Of the unresolved emission, it is likely that about half is
from fainter LMXBs, while about 23\% of the total count rate is due to
interstellar gas with $kT = 0.29$ keV.
The X-ray spectrum of the soft diffuse emission is fit by a thermal
model, and shows clear evidence for soft X-ray lines characteristic
of low density, optically thin gas.
The faint, diffuse soft X-ray emission in the {\it Chandra} image
is very extended, indicating that the ISM gas distribution is much more
extended and less centrally peaked than that of the optical light.
The spatial distribution of the gas is rounder than that of the optical
light, and has a somewhat irregular, L-shaped form.
This may indicate that the gas has been affected by ram pressure
from motions through intergalactic gas.

The X-ray luminosity and spatial distribution of the interstellar
gas in NGC~4697 are inconsistent with a global spherical supersonic wind,
partial wind, or global cooling flow.
The gas may be undergoing subsonic inflation, or a rotationally induced
outflux, or may be affected by ram pressure from intergalactic gas.
There is no apparent relation between the X-ray emitting interstellar
gas and the dust disk or H$\alpha$ emission regions near the center of the
galaxy.

X-ray spectral analysis of the resolved and diffuse emission indicates
that the interstellar gas is the source of the very soft spectral
component seen with {\it ROSAT} in the spectra of this and other
X-ray faint early-type galaxies.
Although the LMXBs produce a large amount of soft X-ray emission and
some of them (the supersoft sources) do have a strong soft X-ray
spectral component, taken together the spectrum of the LMXBs does
not show a significant soft component.
The cumulative LMXB spectrum is well-fit by thermal bremsstrahlung
at $kT = 8.1$ keV.
The soft component in the spectrum is produced mainly by interstellar
gas.

NGC~4697 has an X-ray source located within 1\arcsec\ 
of the optical center with an X-ray luminosity of
$L_X = 8 \times 10^{38}$ ergs s$^{-1}$ (0.3-10 keV).
This source may be due to an AGN and/or one or more LMXBs.
Stellar dynamical measurements indicate that NGC~4697 has a
massive central black hole;
the central X-ray luminosity implies that this BH is emitting at
$\le 4 \times 10^{-8}$ of its Eddington luminosity.

Three of the resolved sources are supersoft sources
(e.g., Kahabka \& van den Heuvel 1997).
These are among the most distant and luminous supersoft sources
known.

Seven of the resolved sources in the outer parts of NGC~4697 (about 20\%)
are coincident with candidate globular clusters.
This implies that globular clusters are a very favorable environment
to form LMXBs;
the same result is true of our Galaxy
(e.g., Hertz \& Grindlay 1983).
We discuss the possibility that all of the LMXBs in NGC~4697 (and
other old stellar systems) were formed in globular clusters.
An {\it HST} observation is needed to determine the globular cluster
population of the inner regions of NGC~4697 where most of the LMXBs
are located.

The X-ray--to--optical luminosity ratio for the LMXBs in NGC~4697 is
$L_X$(LMXB, 0.3--10 keV)$/L_B = 8.1 \times 10^{29}$ ergs s$^{-1}$
$L_{B\odot}^{-1}$.
This is about 35\% higher than the value for the bulge of M31 and
about 13\% higher than the value we found for the S0 galaxy NGC~1553.
This suggests that there are variations
in the X-ray--to--optical ratios early-type galaxies and spiral bulges.
Since much of the X-ray luminosity in NGC~4697 comes from a small
number ($\sim$15) of bright sources, part of this variation may be
due to statistical fluctuations in the number of bright sources or
temporal variations in their individual emission.
If most LMXBs are formed in globular clusters, then their number may
correlate more directly with the population of globulars.
On the other hand, in the range they have in common, the luminosity
function of the bulge of M31 declines more rapidly with increasing
luminosity than in NGC~4697.
This suggests that the populations of LMXBs cannot simply be scaled
from galaxy to galaxy.

The X-ray luminosities (0.3--10 keV) of the resolved LMXBs range from
$\sim$$5 \times 10^{37}$ to $\sim$$2.5 \times 10^{39}$ ergs s$^{-1}$.
The luminosity function has a ``knee'' at
$3.2 \times 10^{38}$ ergs s$^{-1}$, which is approximately the Eddington
luminosity of a 1.4 $M_\odot$ neutron star (NS).
Based on other observations with {\it Chandra},
this knee appears to be a characteristic feature of the luminosity
functions of LMXBs in early-type galaxies.
This knee might provide a standard candle which could be used to
determine distances to galaxies.
This knee may separate accreting NS and
black hole (BH) binaries.
If the brightest sources in NGC~4697 are Eddington limited, they must
contain fairly massive BHs.

These and other {\it Chandra} observations of luminous elliptical galaxies
may represent the first direct detections of neutron stars and stellar mass
black holes in these galaxies.
Our detection of a large population of binaries with NSs and massive BHs
provides perhaps the most direct evidence that this elliptical
galaxy (or its progenitors) once contained a large number of massive
main sequence stars.
The population of LMXBs provides a tool to study the high mass end
of the initial mass function of early-type galaxies, long after the
massive main sequence stars have died.

\acknowledgements
We thank Maxim Markevitch for several extremely helpful communications
concerning the background in the ACIS detector and the readout artifact,
and for very kindly making his blank sky background files and software
available to us.
We are very grateful to Alexey Vikhlinin for providing his software
package for extracting X-ray spectra and constructing the response files
for extended sources.
We also thank Keith Arnaud for advise on extracting spectra from {\it
Chandra.}
We are grateful to JJ Kavelaars for providing his unpublished list
of globular clusters in NGC~4697.
Bill Harris, JJ Kavelaars, and  Arunav Kundu also gave helpful comments
on the globular cluster population of NGC~4697.
We are particularly grateful to Chris Mullis, who took an optical
spectrum of the counterpart of Src.~72 and showed that it was
a background AGN, rather than a globular cluster.
We thank Shri Kulkarni for the interesting suggestion that all or
most LMXBs are formed in globular clusters.
Other useful suggestions were made by Bob O'Connell, Bob Rood, and
Ray White III.
We thank Richard Mushotzky and Lorella Angelini for discussions of their
{\it Chandra} results on X-ray bright ellipticals.
Support for this work was provided by the National Aeronautics and Space
Administration through Chandra Award Numbers
GO0-1019X,
GO0-1141X,
and
GO0-1173X,
issued by
the {\it Chandra} X-ray Observatory Center, which is operated by the Smithsonian
Astrophysical Observatory for and on behalf of NASA under contract
NAS8-39073.
J. A. I. was supported by {\it Chandra} Fellowship grant PF9-10009, awarded
through the {\it Chandra} Science Center.

\clearpage

\clearpage

% ?? figures

\end{document}